\documentclass[12pt,titlepage,epsf,qsymbols]{article}
\usepackage[utf8]{inputenc}
\usepackage[normalem]{ulem}
\usepackage[english]{babel}
\usepackage{tabularx}
\usepackage{array}
\usepackage{graphics}
\usepackage[pdftex]{graphicx}
\usepackage{psfrag}
\usepackage{epsfig}
\usepackage{subfig}
\usepackage{mathtools}
\usepackage{amssymb}

\usepackage{setspace}
\usepackage{rotating}
\usepackage{colortbl}
\usepackage{longtable}
\usepackage{slashed}
\usepackage{braket}
\usepackage{lineno}
\makeatletter
\usepackage{textcomp}
\usepackage[usenames,dvipsnames,table]{xcolor}
\usepackage{relsize}
\usepackage{cite}

\usepackage{float}

\usepackage{bm}
\usepackage{enumerate}
\usepackage{amsmath}
\usepackage{verbatim}
\usepackage[normalem]{ulem}
\usepackage{hyperref}
\hypersetup{colorlinks,citecolor=nicegreen,linkcolor=niceblue}
\hypersetup{colorlinks=true}

\allowdisplaybreaks

\setlongtables

\newcommand{\bea}{\begin{eqnarray}}
\newcommand{\eea}{\end{eqnarray}}
\newcommand{\beq}{\begin{equation}}
{
\newcommand{\eeq}{\end{equation}}
\newcommand{\ec}{\end{center}}
\newcommand{\bc}{\begin{center}}

\newcommand{\gev}{{\rm GeV}}

\newcommand{\pdir}{p\kern -5.2pt\raise 0.2ex\hbox {/}}

\newcommand{\vdir}{v\kern -5.75pt\raise 0.15ex\hbox {/}}
\newcommand{\kdir}{k\kern -5.75pt\raise 0.15ex\hbox {/}}
\newcommand{\epsdir}{\epsilon\kern -5.0pt\raise 0.15ex\hbox {/}}
\newcommand{\bvdir}{\bar{v}\kern -5.75pt\raise 0.15ex\hbox {/}}
\newcommand{\Ddir}{D\kern -7.75pt\raise 0.20ex\hbox {/}}
\newcommand{\Adir}{A\kern -7.75pt\raise 0.20ex\hbox {/}}
\newcommand{\ldir}{l\kern -5.0pt\raise 0.2ex\hbox{/}}
\newcommand{\varepsdir}{\varepsilon\kern -5.5pt\raise 0.15ex\hbox{/}}

\makeatother

\definecolor{niceblue}{rgb}{0.15,0.15,0.6}
\definecolor{nicegreen}{rgb}{0.1,0.5,0.1}
\definecolor{Red}{rgb}{1.,0.,0.}

\definecolor{Green}{rgb}{0.2,.7,0.2}

\begin{document}

\begin{center}

{\Large \bf Testing spin-2 mediator  by  angular observables in $b\rightarrow s \mu^+ \mu^-$  }

\vspace{5mm}

{Svjetlana  Fajfer$^{a,b}$, Bla\v zenka Meli\'c$^{c}$ and Monalisa Patra$^{a,c}$}

\vspace{5mm}

{\it \small
$^{a}$Jo\v{z}ef Stefan Institute, Jamova 39, P. O. Box 3000, 1001 Ljubljana, Slovenia \\
$^{b}$Faculty of Mathematics and Physics, University of Ljubljana, Jadranska 19, 1000 Ljubljana, Slovenia \\
$^{c}$  Institut Rudjer Bo\v skovi\'c, Division of Theoretical Physics, Bijeni\v cka 54, HR-10000, Croatia.
}

\end{center}

\setcounter{page}{1}
\setcounter{footnote}{0}
\setcounter{equation}{0}

\noindent
 We consider effects of spin-2 particle in the $b\rightarrow s \mu^+ \mu^-$ transition assuming that the spin-2 particle couples in a flavour non-universal way to $b$ and $s$ quarks and  in the leptonic sector couples only to the muons, thereby only contributing to the process  $b\rightarrow s \mu^+ \mu^-$. The $B_s - \bar B_s$ transition gives the strong 
constraint on the coupling of the spin-2 mediator and $b$ and $s$ quarks while the  observed discrepancy from the SM prediction for the muon anomalous magnetic moment $ (g-2)_\mu$  serves us to constrain the  $\mu$-coupling to spin-2 particle. 
We find that the spin-2 particle can modify the angular observables in the $B \to K \mu^+ \mu^-$ and $B \to K^* \mu^+ \mu^-$ decays and produce effects which do not exist in the SM. The generated forward-backward asymmetries in these processes can reach $15\%$, while other observables for these decays receive tiny effects.

\renewcommand{\thefootnote}{\arabic{footnote}}

\setcounter{footnote}{0}


\section{Introduction} 
Over the past few years many theoretical and experimental studies of B meson processes were done aimed to test viability of  the Standard Model  (SM). 
On the experimental side precision measurements of many variables were performed, while on the theoretical side complicated B-meson dynamics which requires  precise knowledge of many hadronic parameters obscured precise calculations. 
Although the lattice QCD was able to provide us with very precise determinations of many variables, some information is still lacking.

Such precision on both sides enable us to establish discrepancy  between theoretical prediction and experimental result in the charged current transition $b \to c \tau \nu$ as well as in 
the  flavour changing neutral current transition (FCNC)  of $b\rightarrow s \mu^+\mu^-$. These discrepancies can be searched for in different physical observables which are sensitive to the new physics.
The LHCb collaboration performed measurements of the relevant variables in the FCNC transition $b\rightarrow s \mu^+\mu^-$ within B meson systems.  
The branching ratio of the simplest process of that type $B_s \to \mu^+ \mu^-$ were measured by the both CMS and LHCb collaborations \cite{CMS:2014xfa}
and  the result was found somewhat lower than the predicted one \cite{Bobeth:2013uxa}. The ratio of the branching fractions for the muonic and electronic mode of the $B\to K \ell^+ \ell^-$  decay in the low dilepton invariant mass region $1~\gev^2 \le q^2 \le 6~\gev^2$ \cite{Aaij:2014ora}, known as $R_K$ anomaly, exhibits 2.4$\sigma$ deviations below the SM prediction $R_K^{SM} =1.00(1) $\cite{Hiller:2003js}. Similarly,  the LHCb collaboration  measured  that in the case of $B\to K^* \ell^+ \ell^-$  the experimental result for $R_K^{*}$  disagrees from the SM predictions at the level of (2.2-2.4)$\sigma$  \cite{Aaij:2017vbb}. The deviations from the SM predictions  triggered the questioning of viability of lepton flavour universality (LFU)  \cite{Fajfer:2012jt} and many explanations of the $R_{K^{(*)}}$ anomaly were offered in the literature, mostly assuming some  scenario of New Physics (NP), e.g. \cite{Altmannshofer:2014cfa,Becirevic:2015asa,Datta:2013kja,Hiller:2014yaa,Bauer:2015knc,Glashow:2014iga,Gripaios:2014tna,Greljo:2015mma,Ghosh:2014awa,Crivellin:2015mga,Crivellin:2017zlb,Sierra:2015fma,Varzielas:2015iva,Crivellin:2015era,Celis:2015ara,Ghosh:2017ber,Fajfer:2015ycq,Becirevic:2017jtw,Kamenik:2017tnu,Arnan:2017lxi,Li:2014fea,Kile:2014jea,Bardhan:2017xcc,Guadagnoli:2016erb,Hurth:2017hxg,Crivellin:2017dsk,DAmbrosio:2017wis,Blanke:2018sro,Bhattacharya:2014wla,Feruglio:2016gvd,Buttazzo:2017ixm}.

Apart from the observables $R_{K^{(\ast)}}$,  the angular distributions in $B \rightarrow K^{(*)}\ell^+\ell^-$ decays as a function of dilepton invariant mass squared ($q^2$) are also sensitive to NP effects.
The kinematic distribution of the decay products of $B \rightarrow K \ell^+\ell^-$ at any particular $q^2$ value depends on the angle between the directions of $B$ and of $\ell^-$ in the center of mass frame of the lepton pair. A forward-backward asymmetry constructed at this angle is zero in the SM, therefore a non-zero measurement of the asymmetry will be a clean signal of NP~\cite{Becirevic:2012ec,Bobeth:2007dw}.  Similarly, the full distribution of the decay $B \rightarrow K^*(K\pi) \ell^+\ell^-$ at any particular $q^2$ can be expressed as a triply differential cross-section in three angles namely (a) the angle between the direction of  $K$ in the $K^*$ rest frame and $K^*$ in the $B$ rest frame (b) the angle between $\ell^-$ in the dilepton rest frame and the direction of flight of two leptons in $B$ rest frame and (c) the angle between $\ell^+\ell^-$ and $K\pi$ decay planes in the $B$ rest frame.  Different forward-backward or CP asymmetries can be then constructed by integrating over these angles so as to extract information about NP in detail~\cite{Altmannshofer:2008dz}.

In this study we investigate the effects of spin-2 particle which contributes to the $b\to s \mu^+ \mu^-$ transitions.  
We assume the most general coupling of such massive spin-2 particle, allowing for the flavour violation.  As suggested by  \cite{Grinstein:2012pn} 
the effective Lagrangian study implies that the lowest-order couplings of a spin-2 state with quark fields are quite  similar to the general-relativistic
couplings of the graviton to energy-momentum tensor. The couplings of the spin-2 particle with $b,s$  and with $\mu\mu$ are highly constrained by $B_s-\bar{B}_s$ mixing and $g-2$ of the muon. We show that this scenario therefore  can not explain the experimentally observed $R_K$ anomalies. However, the spin-2 framework induces forward backward asymmetries in $B \to K^{(*)} \mu^+ \mu^-$, which are either zero or very small in the SM and give a clean test of our spin-2 model.  We consider in Sec.~2  the framework of our study introducing interaction of spin-2 particle with fermions of the SM in $b\to s \mu^+ \mu^-$  and then  we constrain model  parameters  using the  $B_s -\bar B_s$  oscillations and muon anomalous magnetic moment $(g-2)_\mu$. Results of our testing of spin-2 mediator in $B\to K^{(*)} \mu^+ \mu^-$ decays are presented in Sec.~3. In Sec.~4 we  discuss numerical results  and shortly summarize the obtained results.

\section{Framework}

The spin-2 particles might appear in the particle physics   due to a number of reasons. For example in lattice QCD  as spin-2 long-lived glueballs \cite{Chen:2005mg}. They also appear  in the modified gravity. Such models contain KK tower of gravitons in theories with large dimensions \cite{ArkaniHamed:1998rs,Antoniadis:1998ig}, wrapped or extra dimensions \cite{Randall:1999ee,Randall:1999vf} . 
We follow most general approach independent of the UV completion as given in Ref. \cite{Grinstein:2012pn}. 
A complex symmetric spin-2  field $G_{\mu \nu}$ having mass M, can be written in the Fierz-Pauli form  \cite{Fierz:1939ix}:
\begin{equation}
\mathcal{L}_{PF} = -\frac{1}{2}G_{\mu \nu}^\dag (\Box + M^2)G^{\mu \nu} +\frac{1}{2} G^{\mu \dag}_{\mu} (\Box + M^2)G^{ \nu}_\nu - G^{\dag}_{\mu \nu} \partial^\mu \partial^\nu G^\rho_\rho +G^\dag_{\mu \nu} \partial^\mu \partial^\rho G^\nu_\rho + h. c. \,.
\label{L1}
\end{equation}
Following Ref.  \cite{Grinstein:2012pn}, we consider an effective Lagrangian for a massive spin-2 boson without any assumption on the origin of the spin-2 field. This means that it might be a non-universal interacting 
gravitational KK mode,  related to RS theory.  Another possibility that $G_{\mu \nu} $ is a bound state of a strongly coupled theory emerging above electroweak scale. 
As pointed out in Ref.  \cite{Choi:1994ax,Han:1998sg,Giudice:1998ck,Degrassi:2008mw,Grinstein:2012pn} an effective Lagrangian describing interactions with the matter fields consists of operators with various dimension.  Suppose, that the scale at which the new interaction emerges is well above electro-weak scale. Then the operators in the effective Lagrangian have to be  suppressed by the appropriate dimension of this scale. 
The dimension 4 Lagrangian with no derivative on the fermionic fields will have a form of
\begin{equation}
\mathcal{L}_{4} \subset - G^\mu_\mu  \lambda_{ij} \bar{\psi}_i \psi_j   + h. c. \,.
\label{L2}
\end{equation}
Such interaction is, in its form, similar to the interaction of a  scalar field, 
although the correlator for spin-2 particle is different from the spin-0 one.  In our study we would not consider these effects since they can be reduced to the effects of the scalar/pseudoscalar  operators in the effective Lagrangian 
describing the $b \to s \mu^+\mu^-$ transitions as performed in Ref. \cite{Becirevic:2012ec}. According to analysis of Ref. \cite{Capdevila:2017bsm}  such operators, however, cannot explain $R_{K^{(*)}}$ anomalies. 

The most general coupling of the spin-2 particle with the fermions at 
the one derivative order is given by the dimension five  Lagrangian ~\cite{Choi:1994ax,Han:1998sg,Giudice:1998ck,Degrassi:2008mw,Grinstein:2012pn}
\begin{eqnarray}\label{eq:lagrangian}
\mathcal{L}_5&=& -\frac{i}{4\Lambda}\left\lbrace a^L_{ij} \left[\bar{\psi}_{i}\left(\gamma_\mu \partial_\nu 
 + \gamma_\nu \partial_\mu \right) P_L \psi_j\right] \right. \nonumber \\
  &&\left. \qquad +b^L_{ij} \left[ \left(\partial_\nu \bar{\psi}_i\gamma_\mu 
 +  \partial_\mu \bar{\psi}_i \gamma_\nu  \right) P_L \psi_j\right]\right\rbrace G^{\mu\nu} + (L \to R)
\end{eqnarray}
where the constants $a_{ij}, b_{ij}$ will be taken as real, and in addition we take $a_{ij} = a_{ji}, b_{ij} = b_{ji}$. 
Although we consider a general spin-2 particle, similarly as in \cite{Grinstein:2012pn}, we will restrict the couplings by imposing 
\begin{eqnarray}\label{eq:const_constr}
a^{L,R}_{ij} = -b^{L,R}_{ij}, 
\end{eqnarray}  
which appear to be valid for spin-2 particle couplings with fermions in the gravitation theory (for e.g. in the context of the RS models with the SM fermions propagating in the bulk.)~\cite{Grossman:1999ra,Gherghetta:2000qt,Agashe:2004cp,Agashe:2006wa,Agashe:2007zd}.  The  couplings $a^{L,R}_{ij}$  are going to be  arbitrary parameters, although, due to the V-A nature of the charged weak current couplings, one expects that $a^{R}_{ij}$ couplings are highly suppressed.  We also consider the perturbative limits for the couplings of spin-2 with the quarks and the leptons.   Since we only consider the consequences of the spin-2 field on the $b\rightarrow s \mu \mu$ transitions, we do not specify the UV completion of the theory.   A detailed discussion on the range of validity of this low-energy effective theory in the context of theories with a massive general relativity in 4D can be found in~\cite{Grinstein:2012pn}. 

There are also additional terms in the Lagrangian of the form
\begin{eqnarray}
\mathcal{L'}_5 &=& -\frac{1}{2\Lambda}  \eta_{\mu \nu} \left\lbrace c_{ij}^L   \bar{\Psi}_{i}  i \gamma_\rho \partial^\rho  \Psi_j    
  \right\rbrace  G^{\mu\nu}  
+ (L \to R), 
\end{eqnarray}
which might contribute too. After using equation of motion  for fermion fields, these terms will again produce scalar and pseudoscalar operator contributions being highly suppressed  Ref. \cite{Degrassi:2008mw,Becirevic:2012ec}.
\footnote{In \cite{Degrassi:2008mw} the authors have written the exact structure of SM fermion interactions with gravitons at the one-loop level, which could be then promoted to the most generalized interaction Lagrangian of fermions with a spin-2 field. However, since we are here interested only into leading effects, we will work only with the leading part of the Lagrangian (\ref{eq:lagrangian}) (with constraint (\ref{eq:const_constr}) taken into account).}

\begin{figure}[htb]
\begin{center}
 \includegraphics[width=9cm, height=5.5cm]{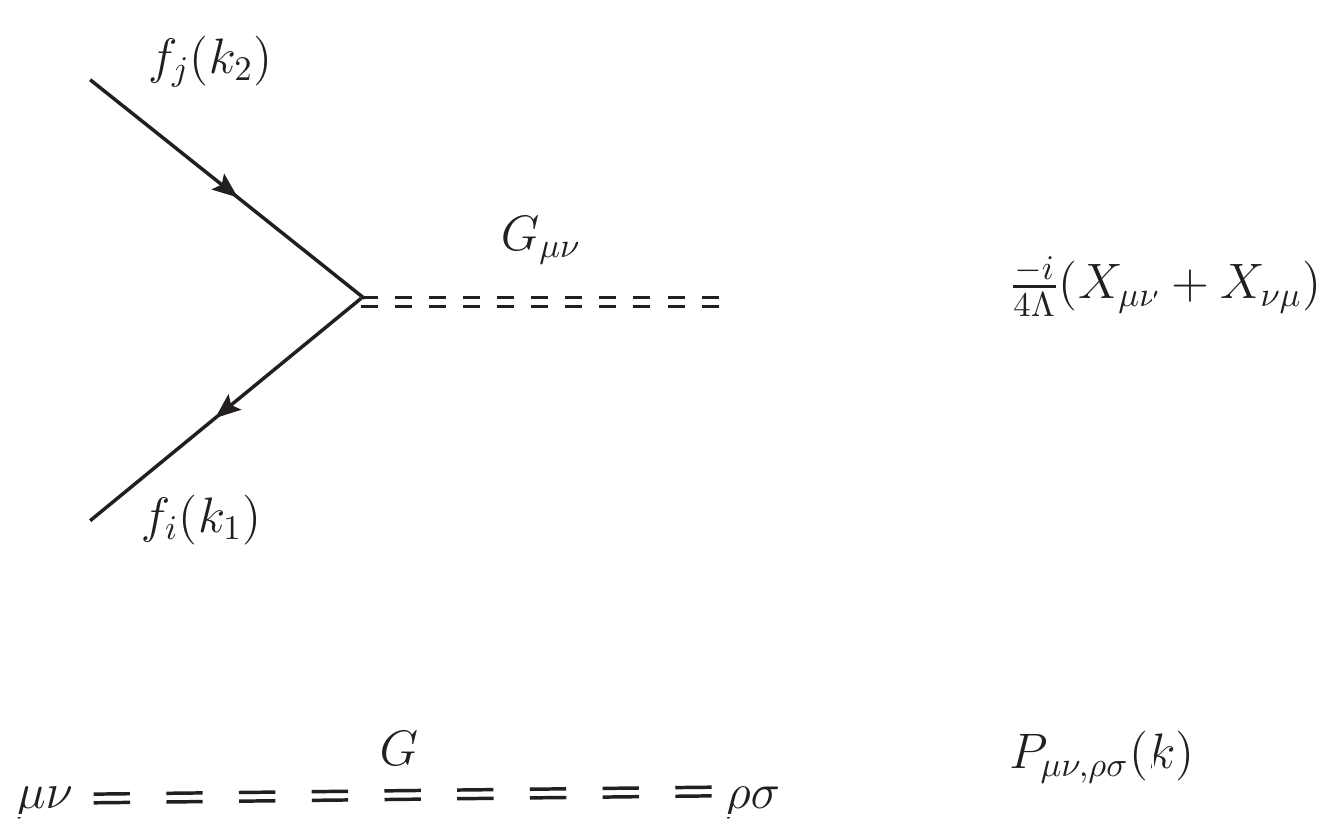}
\end{center}
 \caption{Three-point vertex and the propagator used for our calculation, with 
 $X_{\mu\nu}$ and $P_{\mu\nu,\rho\sigma}$ defined in Eq.~\ref{eq:vertex}. 
 }
 \label{fig:feynman}
\end{figure}

Within framework described above, the Feynman rules for the spin-2 interactions with the SM fermions in the de Donder gauge~\cite{Han:1998sg} are given in Fig.~\ref{fig:feynman}, with the symbols defined as:
\begin{eqnarray}
 X_{\mu\nu} &=& a^L_{ij}\gamma_\mu (k_{1\nu}+k_{2\nu})P_L + a^R_{ij}\gamma_\mu (k_{1\nu}+k_{2\nu})P_R \nonumber \\
 P_{\mu\nu,\rho\sigma}(q^2) &=& \frac{i}{q^2-m_G^2}
 \left[\left(\eta_{\mu\rho}-\frac{q_\mu q_\rho}{m_G^2}\right)\left(\eta_{\nu\sigma}-\frac{q_\nu q_\sigma}{m_G^2}\right)
 +\left(\eta_{\nu\rho}-\frac{q_\nu q_\rho}{m_G^2}\right)\left(\eta_{\mu\sigma}-\frac{q_\mu q_\sigma}{m_G^2}\right), 
 \right.\nonumber \\
 &&\left. -\frac{2}{3}\left(\eta_{\mu\nu}-\frac{q_\mu q_\nu}{m_G^2}\right)\left(\eta_{\rho\sigma}-\frac{q_\rho q_\sigma}{m_G^2}\right)\right]. 
 \label{eq:vertex}
\end{eqnarray}
Provided the momenta transfer $q^2 \ll m_G^2$, the non-local interaction can be approximated to a local interaction,
using Taylor's expansion 
\begin{eqnarray}
 \frac{1}{q^2-m_G^2}=-\frac{1}{m_G^2}\left[\frac{1}{1-q^2/m_G^2}\right]=-\frac{1}{m_G^2}
 \left[1+\frac{q^2}{m_G^2}+\frac{q^4}{m_G^4}+\cdots\right]
\end{eqnarray}

On the other hand, we are interested in the most general effective Hamiltonian describing the $b\rightarrow s\mu^+\mu^-$ transitions at low energy.  Such process  at scale $\mu = \mu_b = 4.8$  GeV are governed by dimension 6 effective Hamiltonian:  
\begin{eqnarray}\label{eq:Leff}
\mathcal{H}_{\mathrm{eff}} =-\frac{4 G_F}{\sqrt{2}}V_{tb}V^*_{ts}  \bigg( \sum_{i}  C_i (\mu) \mathcal{O}_i (\mu)
+C'_i(\mu) \mathcal{O}'_i (\mu)  +   \sum_{j;h.d.} C^{h.d.}_j \mathcal{O}^{h.d.}_j \bigg ) + {\rm h.c},
\end{eqnarray}
where the operators are
\begin{eqnarray}
\mathcal{O}_7 &=&\frac{e^2}{g^2}m_b (\bar{s}\sigma_\mu P_R b) F^{\mu\nu} , \qquad \qquad 
\mathcal{O}'_7 =\frac{e^2}{g^2}m_b (\bar{s}\sigma_\mu P_L b) F^{\mu\nu} ,\nonumber \\
\mathcal{O}_9 &=& \frac{e^2}{16\pi^2}(\bar{s}\gamma_\mu P_L b)(\bar{\ell} \gamma^\mu \ell), \qquad \qquad \qquad 
\mathcal{O}'_9 = \frac{e^2}{16\pi^2}(\bar{s}\gamma_\mu P_R b)(\bar{\ell} \gamma^\mu \ell), \nonumber \\
\mathcal{O}_{10} &=& \frac{e^2}{16\pi^2}(\bar{s}\gamma_\mu P_L b)(\bar{\ell} \gamma^\mu \gamma_5 \ell), \qquad \quad \qquad 
\mathcal{O}'_{10} = \frac{e^2}{16\pi^2}(\bar{s}\gamma_\mu P_R b)(\bar{\ell} \gamma^\mu \gamma_5 \ell), \nonumber \\ 
\mathcal{O}_S &=& \frac{e^2}{16\pi^2}(\bar{s}P_R b)(\bar{\ell} \ell), \qquad \qquad \qquad \quad
\mathcal{O}'_S = \frac{e^2}{16\pi^2}(\bar{s}P_L b)(\bar{\ell} \ell), \nonumber \\
\mathcal{O}_P &=& \frac{e^2}{16\pi^2}(\bar{s}P_R b)(\bar{\ell}\gamma_5 \ell), \qquad \qquad \quad \quad
\mathcal{O}'_P = \frac{e^2}{16\pi^2}(\bar{s}P_L b)(\bar{\ell}\gamma_5 \ell) ,\nonumber \\
\mathcal{O}_T &=& \frac{e^2}{16\pi^2}(\bar{s}\sigma_{\mu\nu} P_L b)(\bar{\ell} \sigma^{\mu\nu} \ell),\qquad \quad \quad
\mathcal{O}_{T5} = \frac{e^2}{16\pi^2}(\bar{s}\sigma_{\mu\nu} P_L b)(\bar{\ell} \sigma^{\mu\nu} \gamma_5 \ell).\qquad \quad \qquad 
\end{eqnarray} 
Note that the operators $\mathcal{O}_i$, $i=7$, $9$, $10$ appear in the SM physics. 
Usually NP also generates operators of the dimension 6, although it is sometimes necessary to enlarge the base of operator to the higher dimensions, like for example, in the case of Two Higgs Doublet Model \cite{Arnan:2017lxi,Li:2014fea} where the operator basis has to be enlarged by the dimension 7 operators. 
In the next section we will show in our model the enlargement of the operator basis by operators of the dimension 8 induced by the spin-2 particle mediation. 

\subsection{Spin-2 mediator in the $b\rightarrow s \mu^+ \mu^-$ transition}

\begin{figure}[htb]
\centering
 \includegraphics[width=9cm, height=4cm]{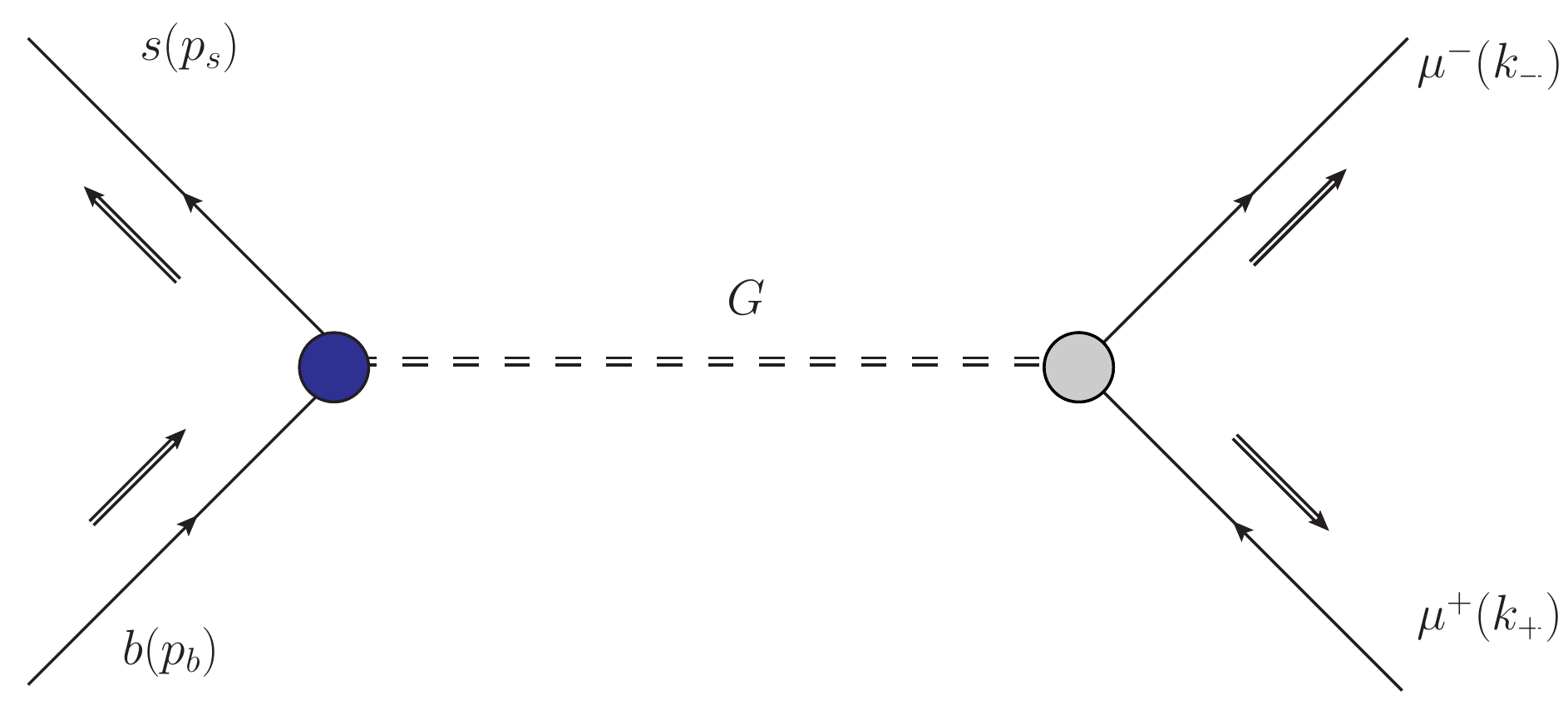}
 \caption{Feynman diagrams for the $b\rightarrow s \mu^+ \mu^-$ transition with the spin-2 mediation.}
 \label{fig:diagram}
\end{figure}

We first consider the contributions of the spin-2 particle  to the  $b\rightarrow s \mu^+ \mu^-$ amplitude, as presented in Fig.~\ref{fig:diagram}. 
By such choice we assume that this fermion spin-2 interaction is flavour non-universal and that only relevant couplings are $bs$ to spin-2 particle  and $\mu \mu$ to spin-2 particle couplings. In the calculation of the $b(p_b) \rightarrow s(p_s) \mu^-(k_-) \mu^+(k_+)$ amplitude, we use the following notation
$q = p_b - p_s = k_+ + k_-$, $p = p_b +p_s$ and $k = k_+ - k_-$. 

%
%
The amplitude for the $b\rightarrow s \mu^+ \mu^-$ transition is then of the form:
\begin{eqnarray}
 i\mathcal{M} &=& \left(\frac{-i}{4\Lambda}\right)^2 a^i_{sb} a^j_{\mu \mu} 
 \bigg\lbrace \bar{u}_{s}(p_s)\left[\gamma^\mu (p_s^\nu+p_b^\nu) + 
 \gamma^\nu (p_s^\mu+p_b^\mu)\right]P_i u_b(p_b)\bigg\rbrace P_{\mu\nu,\rho\sigma} \nonumber \\
 && \qquad \quad   \bigg\lbrace \bar{u}_{\ell}(k_-)\left[\gamma^\rho (k_-^\sigma-k_+^\sigma) + 
 \gamma^\sigma (k_-^\rho-k_+^\rho)\right] P_j v_\ell(k_+)\bigg\rbrace \nonumber \\
 &=& i \frac{a^i_{sb}a^j_{\mu \mu} }{16 \Lambda^2 m_G^2} 
 \bigg\lbrace \bar{u}_{s}(p_s)\left[\gamma^\mu (p_s^\nu+p_b^\nu) + 
 \gamma^\nu (p_s^\mu+p_b^\mu)\right]P_i u_b(p_b)\bigg\rbrace  T_{\mu\nu}^j,
\end{eqnarray}
where $T_{\mu\nu}$ is obtained through the contraction of the propagator with the leptonic current, 
\begin{eqnarray}
T_{\mu\nu}^L &=& -\frac{4}{3} m_{\ell} \bigg[\bar{u}_{\ell}(k_-)P_L v_\ell(k_+) 
+ \bar{u}_{\ell}(k_-)P_R v_\ell(k_+)\bigg] g_{\mu\nu}
+2(k_{-\mu}-k_{+\mu}) \bigg[\bar{u}_{\ell}(k_-)\gamma_\nu P_L v_\ell(k_+)\bigg]\nonumber \\
&& +2 (k_{-\nu}-k_{+\nu})\bigg[\bar{u}_{\ell}(k_-)\gamma_\mu P_L v_\ell(k_+)\bigg],\nonumber \\
T_{\mu\nu}^R &=&  T_{\mu\nu}^L \, ({\rm with \; L \leftrightarrow R}). 
\end{eqnarray}

After multiple use of the equation of motions,  we determine  contributions to the scalar operators of the dim-8 
\begin{equation}
 \mathcal{O}_{S}^{(q,8)} = m_\ell  m_q \frac{e^2}{16\pi^2}(\bar{s}P_R b)(\bar{\mu} \mu),\quad
\mathcal{O}_{S^{\prime}}^{(q,8)}  = m _\ell m_q \frac{e^2}{16\pi^2}(\bar{s}P_L b)(\bar{\mu} \mu), 
\label{scalars}
\end{equation}
with $q=s,b$. The product $m_\ell m_s$ being small can be safely neglected. We therefore consider the operators $\mathcal{O}_{S}^{(b,8)},~\mathcal{O}_{S^{\prime}}^{(b,8)}$, henceforth referred to as $\mathcal{O}_{S}^{(8)},~\mathcal{O}_{S^{\prime}}^{(8)}$. The appropriate Wilson coefficients for the operators proportional to $m_\ell m_b$ are:
\begin{eqnarray}
\label{eq:SMspin2}
C_{S}^{(8)} &=& \frac{4}{3} C_G\,  a^L_{sb} \left[a^L_{\mu \mu} +a^R_{\mu \mu}\right] ,\nonumber \\
C_{S'}^{(8)} &=& \frac{4}{3} C_G \, a^R_{sb}  \left[a^L_{\mu \mu} +a^R_{\mu \mu}\right] , 
\end{eqnarray}
where we used:
\begin{equation}
C_G =\frac{16 \pi^2 v^2}{e^2} \frac{1}{V_{tb}V^*_{ts}} \frac{1}{16 \Lambda^2 m_G^2} .
\label{CG}
\end{equation}
Here the dimension 8 operators are proportional to  the product of $m_b m_{\mu}$ with the dimension 6 operators 
${\cal O}_S$ and ${\cal O}_{S^\prime}$. However, there are new additional operators of the dimension 8: 
\begin{eqnarray}
\label{eq:newoperators}
 \mathcal{O}^{(8)}_{i}  &=&
(\bar{s} \gamma^\mu i \overset{\leftrightarrow}{\partial^\nu} P_i b)
(\bar{\ell} \gamma_\mu i \overset{\leftrightarrow}{\partial}_\nu \ell),\nonumber \\
\mathcal{O}^{(8)}_{i5}  &=& (\bar{s} \gamma^\mu i \overset{\leftrightarrow}{\partial^\nu} P_i b)
(\bar{\ell} \gamma_\mu i \overset{\leftrightarrow}{\partial}_\nu \gamma_5 \ell), \nonumber \\
 \mathcal{O}^{(8)}_{ij} &=&  (\bar{s} \gamma^\nu i \overset{\leftrightarrow}{\partial^\mu} P_i b)
(\bar{\ell} \gamma^\mu i \overset{\leftrightarrow}{\partial^\nu} P_j \ell) ,
\end{eqnarray}
with the Wilson coefficients:
\begin{eqnarray}
\label{eq:newoperators}
{C}^{(8)}_i = -C_G\,a^{i}_{sb} (a^{L}_{\mu \mu}+a^{R}_{\mu \mu}),\quad 
{C}^{(8)}_{i5} = -C_G\, a^{i}_{sb} (a^{R}_{\mu \mu}-a^{L}_{\mu \mu}),\quad 
{C}^{(8)}_{ij}= -2 C_G\,  a^i_{sb} a^j_{\mu \mu}.
\end{eqnarray}
By using the Fierz rearrangement and by applying the field equations, one can show that 
\begin{eqnarray}
\int d^4 x \langle l^+(k_+) l^-(k_-) s(p_s) | {\cal O}_i^{(8)} (x) | b(p_b) \rangle \rightarrow (k\cdot p)  
\Big[\bar{u}_s(p_s) \gamma^\mu P_i u_b(p_b)\Big]
\Big[\bar{u}_{\ell}(k_-)\gamma_\mu  v_\ell(k_+)\Big] \nonumber \\
\int d^4 x \langle l^+(k_+) l^-(k_-) s(p_s) | {\cal O}_{i5}^{(8)} (x) | b(p_b) \rangle \rightarrow (k\cdot p)  
\Big[\bar{u}_s(p_s) \gamma^\mu P_i u_b(p_b)\Big]
\Big[\bar{u}_{\ell}(k_-)\gamma_\mu \gamma_5  v_\ell(k_+)\Big] 
\nonumber \\
\int d^4 x \langle l^+(k_+) l^-(k_-) s(p_s) | {\cal O}_{ij}^{(8)} (x) | b(p_b) \rangle \rightarrow (k^\mu p^\nu)  
\Big[\bar{u}_s(p_s) \gamma_\mu P_i u_b(p_b)\Big]
\Big[\bar{u}_{\ell}(k_-)\gamma_\nu P_j v_\ell(k_+)\Big] 
\label{opp}
\end{eqnarray}
from which we can see that effectively the dim-8 operators ${\cal O}_i^{(8)}$ and ${\cal O}_{i5}^{(8)}$ have the same Lorentz structure as the dim-6 operators which matrix elements can be calculated. Namely we notice that
${\cal O}_i^{(8)}  \sim  {\cal O}_9, {\cal O'}_9 $ and 
${\cal O}_{i5}^{(8)} \sim {\cal O}_{10}, {\cal O'}_{10}$. 
The results of the global fit analysis \cite{Capdevila:2017bsm}  suggests that 
the B-physics anomalies can be explained by the NP for the Wilson coefficients which satisfy  the relation $C_9 \simeq -C_{10} $. This indicate that the new interactions should be of $V-A$ type. 
In order to follow such $V-A$ frame of NP,  we   impose condition: \begin{eqnarray}
a^R_{sb}, a^R_{\mu \mu} = 0  \quad \rightarrow \quad C^{(8)}_i = -C^{(8)}_{i5} .
\end{eqnarray}
The operator $\mathcal{O}^{(8)}_{LL}$,
as can be seen from (\ref{opp}), is a genuine operator of dim-8 that  cannot be reduced to the standard dim-6 operators. This means that we have three independent  operators of the type ${\mathcal O}^{(8)}_{L}$ , ${\cal O}^{(8)}_{L5}$ 
and ${\mathcal O}^{(8)}_{LL}$ and two scalar operators ${\mathcal O}_S^{(b,8)}\equiv{\mathcal O}_S^{(8)}$ and ${\mathcal O}_{S^\prime}^{(b,8)}\equiv{\mathcal O}_{S^\prime}^{(8)}$  (Eq.~\ref{scalars}) with the appropriate Wilson coefficients: 
\begin{eqnarray}
C^{(8)}_L &=& -C_G\, a^{L}_{sb} a^{L}_{\mu \mu}, \quad C^{(8)}_{L5} = C_G \,a^{L}_{sb} a^{L}_{\mu \mu},\quad C^{(8)}_{LL} = -2 C_G \,a^{L}_{sb} a^{L}_{\mu \mu},  
\nonumber \\
C^{(8)}_S &=& \frac{4}{3} C_G\, a^{L}_{sb} a^{L}_{\mu \mu}, \quad C^{(8)}_{S'} = 0.
\end{eqnarray}
In the following we will consider effects of this operator basis.  The new contribution can appear in the decay amplitudes for  $B_s \to \mu^+ \mu^-$, $B\to K \mu^+ \mu^-$ and $B\to K^*  \mu^+ \mu^-$.

\subsection{Constraints on spin-2 couplings 
}

Our model independent analysis aimed to determine spin-2 mediator effects in $b\rightarrow s \mu^+ \mu^-$  requires knowledge of the $bs-$ spin-2 coupling  $a^L_{sb}$ and $\mu \mu-$ spin-2 coupling $a^L_{\mu\mu}$. First we discuss the  determination of $a^L_{sb}$.

\subsection{$B_s -\bar{B}_s$ mixing}
The coupling of the $b$ and $s$ quarks with the  spin-2 particle can be extracted from the $B_s -\bar{B}_s$  mixing observables. 
\begin{figure}[htb]
\centering
 \includegraphics[width=16cm, height=3.2cm]{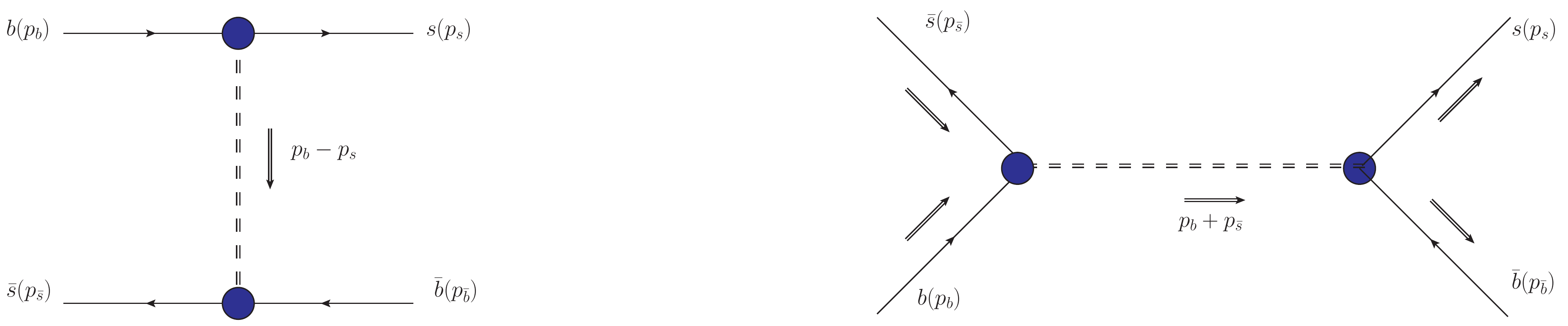}
 \caption{A tree level contribution to $B_s- \bar{B}_s$ mixing.}
 \label{fig:bbar}
\end{figure}
The spin-2 tree-level contribution to the  $B_s-\bar{B}_s$ transition is presented in Fig.~\ref{fig:bbar}. 
The most general effective Hamiltonian of dim-6 for $\Delta B= 2$ process \cite{Becirevic:2001jj,Buras:2001ra,Buras:2013td,Herrlich:1996vf,Lenz:2006hd} 
can be written as:
\begin{equation}
H_{eff}^{\Delta B=2}=\sum_{i=1}^5 C_i \mathcal{Q}_i + \sum_{i=1}^3 \tilde{C}_i \tilde{\mathcal{Q}}_i.  
\end{equation}
The full set of operators $\mathcal{Q}_i$ and $ \tilde{\mathcal{Q}}_i$  are presented in  \cite{Becirevic:2001jj,Buras:2001ra,Buras:2013td,Herrlich:1996vf,Lenz:2006hd,DiLuzio:2017fdq}. The current experimental result for  $\Delta m_{B_s} =(1.1688 \pm 0.0014)\times 10^{-11}$ GeV \cite{Patrignani:2016xqp} is consistent with the SM result ($1.10083^{+0.054} _{-0.038}$) $\times 10^{-11}$ GeV \cite{Lenz:2011ti}.

In the case of spin-2 mediator the operators are of dimension 8. These operators can be written as the product of 
quark masses and  the dimension-6 operators $\mathcal{Q}_{i},~\tilde{\mathcal{Q}}_{i}$, with $i$ = 2,3,4,5.
\begin{eqnarray}\label{eq:bbar}
\mathcal{Q}^{(8)}_2 &=&m_s^2 (\bar{s}_R^\alpha b_L^\alpha) (\bar{s}_R^\beta  b_L^\beta),\quad \quad  \quad \quad\tilde{\mathcal{Q}}^{(8)}_2 = m_b^2  (\bar{s}_L^\alpha b_R^\alpha) (\bar{s}_L^\beta  b_R^\beta),\nonumber \\
\mathcal{Q}^{(8)}_3 &=&m_s^2 (\bar{s}_R^\alpha b_L^\beta) (\bar{s}_R^\beta  b_L^\alpha),\quad \quad \quad \quad  \tilde{\mathcal{Q}}^{(8)}_3 =m_b^2 (\bar{s}_L^\alpha b_R^\beta) (\bar{s}_L^\beta  b_R^\alpha),\nonumber \\
\mathcal{Q}^{(8)}_4 &=&m_b m_s (\bar{s}_R^\alpha b_L^\alpha) (\bar{s}_L^\beta  b_R^\beta), \quad \quad \quad 
\mathcal{Q}^{(8)}_5 =m_b m_s (\bar{s}_R^\alpha b_L^\beta) (\bar{s}_L^\beta  b_R^\alpha),
\end{eqnarray}
There are also two additional operators: \begin{eqnarray}
\mathcal{Q}^{(8)}_6 &=&  
(\bar{s}^\alpha \gamma^\mu i \overset{\leftrightarrow}{\partial}_\nu P_L b^\alpha)
(\bar{s}^\beta \gamma^\mu i \overset{\leftrightarrow}{\partial}_\nu P_L b^\beta),  \quad
\mathcal{Q}^{(8)}_7 =  (\bar{s}^\alpha \gamma^\nu i \overset{\leftrightarrow}{\partial}_\mu P_L b^\alpha)
(\bar{s}^\beta \gamma^\mu i \overset{\leftrightarrow}{\partial}_\nu P_L b^\beta).
\end{eqnarray}
 By the use of equations of motion the operator $\mathcal{Q}^{(8)}_6$  can be reduced to $m_b^2 \mathcal{Q}_1$ with $\mathcal{Q}_1 = (\bar{s}_L^\alpha \gamma^\mu b_L^\alpha) (\bar{s}_L^\beta \gamma^\mu b_L^\beta)$. 
 In our numerical study the operators proportional to $m_s^2$ and $m_b m_s$  can be safely neglected. 
The corresponding Wilson coefficients of the  operators are 
\begin{eqnarray}
 &&C^{(8)}_2= - C^{(8)}_3  =\frac{1}{2} C^{(8)}_4  =  \tilde{C}^{(8)}_2= -\tilde{C}^{(8)}_3  = \frac{1}{2} C^{(8)}_5=-\frac{(a^L_{sb})^2}{ \Lambda^2 m_G^2} \left(\frac{8}{3}\right), \, \nonumber \\
&&C^{(8)}_6 = C^{(8)}_7 = 4\frac{(a^L_{sb})^2}{16 \Lambda^2 m_G^2} 
\label{Bs-coef} 
\end{eqnarray}
The hadronic matrix elements of the operators $m_b^2 (\mathcal{Q}_1, \tilde{\mathcal{Q}}_{2,3})$ between the neutral $B_s$ mesons and the corresponding Wilson coefficients at the scale  $\Lambda$ evolved to the hadronic scale with the anomalous dimension matrices can be found in \cite{Becirevic:2001jj}. We added the contribution of the spin-2 mediator to the mass difference of the $B_s -\bar{B}_s$ system to the SM contribution and write it schematically as: 
\begin{equation}
\Delta M_s = (\Delta M_s)_{SM}+ (\Delta M_s)_{m_b^2\mathcal{Q}_1} +(\Delta M_s)_{m_b^2\tilde{\mathcal{Q}}_2}+(\Delta M_s)_{m_b^2\tilde{\mathcal{Q}}_3}.
\label{MBs}
\end{equation}
The results with spin-2 contributions are given in Appendix A. Note that the  spin-2 contributions  contain the common  proportionality factor $\frac{(a^L_{sb})^2}{\Lambda^2 m_G^2}$. 
Following the calculation of \cite{Becirevic:2001jj}, we determine bound on the $b\, s -$ spin-2 particle coupling:
\begin{equation}
\left(\frac{a^L_{sb}}{\Lambda m_G}\right)^2 < 1.4~\times 10^{-12}\, {\rm GeV}^{-4}.
\label{BBbarmix}
\end{equation}
One can find straightforwardly  that for $\Lambda \simeq 1$ TeV and $m_G \simeq 500$ GeV, $a^L_{sb} \approx 0.5$. 
\begin{figure}[htb]
 \centering
\includegraphics[width=7cm, height=6cm]{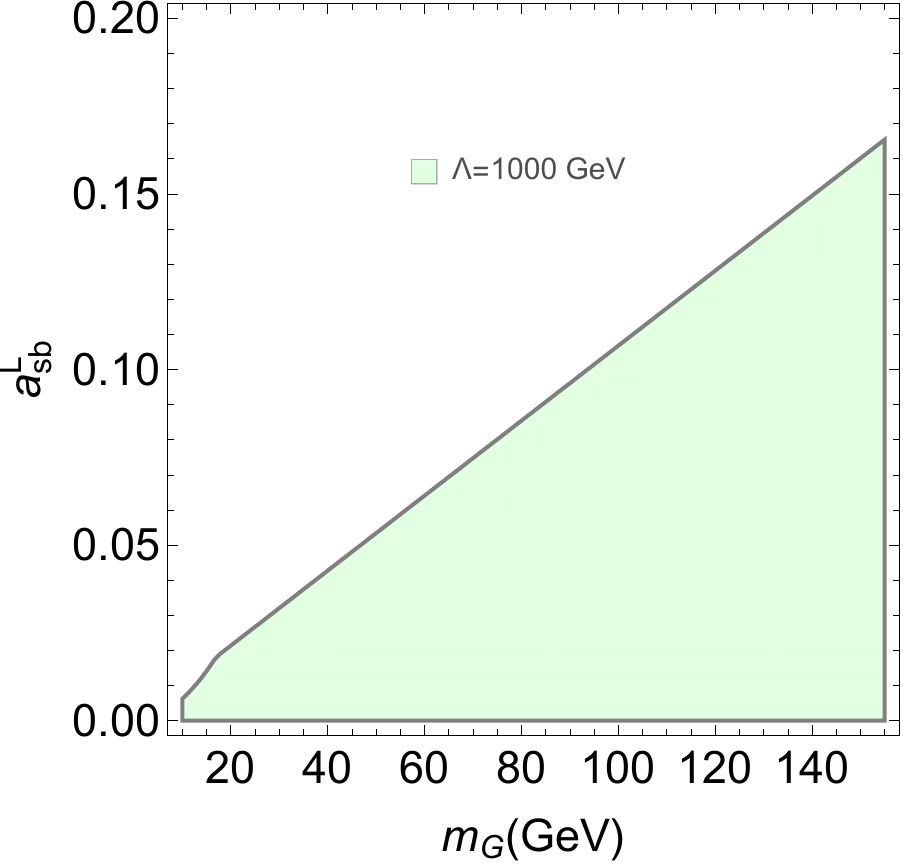}
 \caption{Combined bound  from  $B_s-\bar{B}_s$ mixing and the top decay width on $a^L_{sb}$ as a function of $m_G$.}
\label{fig:boundaLsb}
\end{figure}
We point out that due to  the $SU(2)_L$ symmetry one can relate the coupling of top quark, charm quark and spin-2 particle to the coupling of bottom, strange quarks and spin-2 mediator. The top quark decay channel $t\rightarrow c ~G_{\mu\nu}$, is allowed when the mass of the spin-2 particle is less than the top quark mass.
Using  $a^L_{sb} \simeq a^L_{tc}$, the decay width of this mode is given by 
\begin{equation}
 \Gamma_t^{NP} =\frac{|a^L_{sb}|^2m_t^7}{192 \pi \Lambda^2 m_G^4}\left(1-\frac{m_G^2}{m_t^2}\right)^4
 \left(2+3\frac{m_G^2}{m_t^2}\right).
 \label{GammaTop}
\end{equation}
The SM decay width for $m_t$ =173.3 GeV is 1.35 GeV, whereas the measured decay width is
$1.41^{+0.19}_{-0.15}$ GeV. The top decay width might give a strong bound on $a^L_{sb}$
for  $m_G < m_t$. We show in Fig.~\ref{fig:boundaLsb} the combined bound from  the $B_s-\bar{B}_s$ mixing (\ref{BBbarmix}) and the top decay width (\ref{GammaTop}) on $a^L_{sb}$ as a function of $m_G$ for $\Lambda$ = 1000 GeV.  We consider $\Lambda \simeq$ 1000 GeV along with $m_G$ in the mass range 10-150 GeV, so that the current LHC data does not constrain the parameter space considered here. In case of masses above 150 GeV, the decay channel to massive gauge bosons or to a pair of Higgs boson opens up. These channels have been explored in the LHC and limits are obtained on the parameter space of the respective models.   Similarly, in the case of spin-2 particle mass below 10 GeV, the upsilon decay to dimuon final state will open and this process has been studied in detail. This in turn leads to strong constraints on spin-2 mass below 10 GeV.

\subsection{Constraints from $(g-2)_\mu$ }

\begin{figure}[htb]
 \centering
  \includegraphics[width=5cm, height=4.5cm]{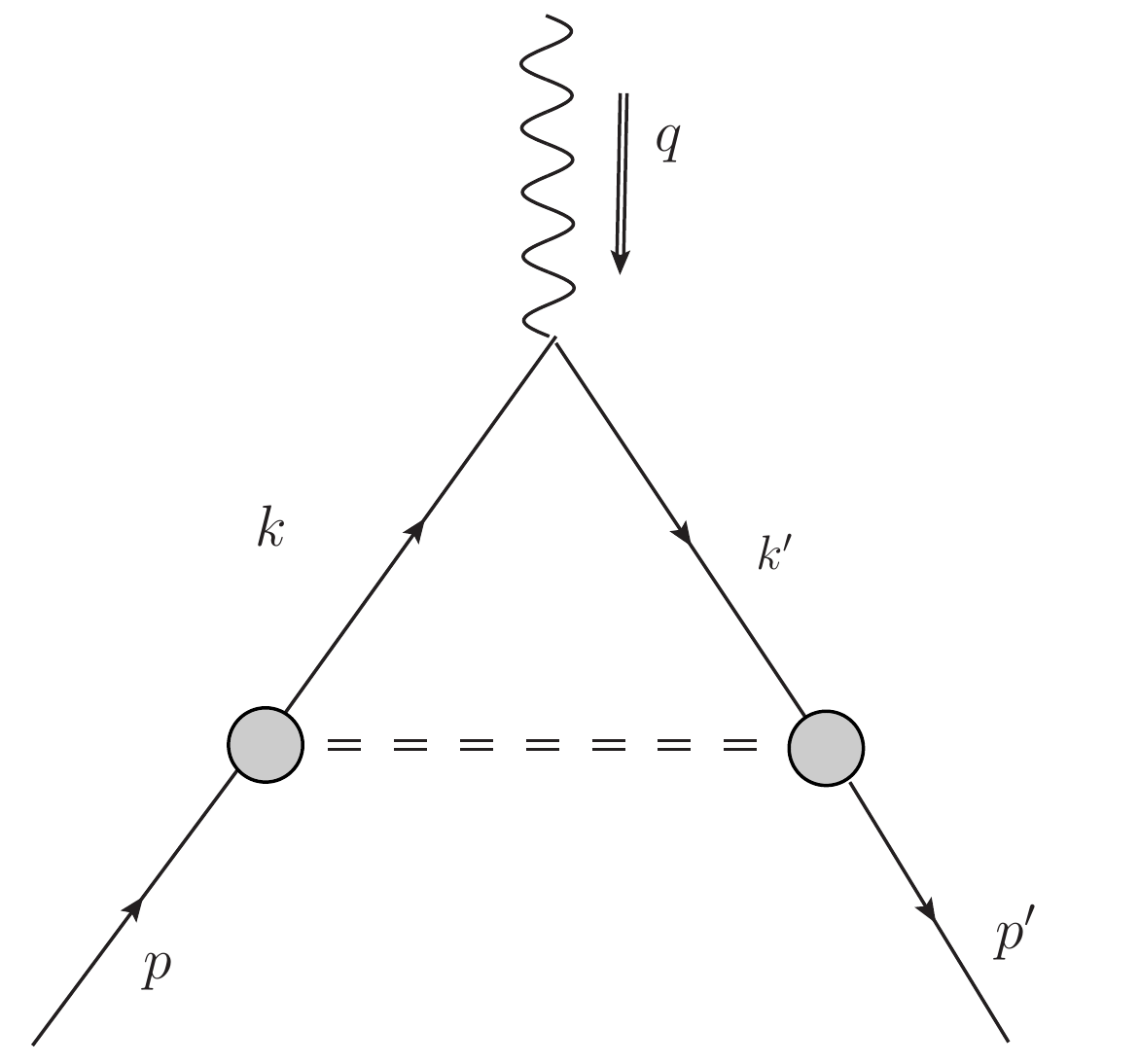}
  \caption{The lowest order contribution of spin-2 particle.}
  \label{fig:gminus2}
 \end{figure}

 The experimental and theoretical results for the muon anomalous magnetic moment  $a_\mu= 1/2(g-2)$  differ. The  recent update of the well established difference is   $\Delta a_\mu= a_\mu^{exp}- a_\mu^{th} = (3.11\pm 0.77)\times 10^{-9}$  \cite{Jegerlehner:2017lbd}. 
 In the presence of spin-2 particle the lowest order contribution to the muon anomalous magnetic moment is shown in 
 Fig.~\ref{fig:gminus2}. 
We calculate the spin-2 mediator contributions to  the muon anomalous magnetic moment $(g-2)_\mu$  and find:
\begin{align}
 &\left(\frac{a^L_{\mu\mu}}{4\pi}\right)^2\frac{2m_\mu}{(4\Lambda)^2} 
 \int_0^1 dx~dy~dz~\delta(x+y+z-1)~\bar{u}(p')\left[i \frac{\sigma^{\mu\nu}q_\nu}{2m_\mu} 
 \bigg(-\frac{2m_\mu^3}{3\Delta}\left(6z^4+26 z^3 \right. \right. \nonumber \\
 &\left.\left.+ 29 z^2 -5z-11\right) -\frac{4m_\mu}{3} \left\lbrace \mathrm{log}\bigg[\frac{\Lambda_N^2+\Delta}{\Delta}\bigg]
 +\frac{\Delta}{\Lambda_N^2+\Delta}-1\right\rbrace
 (6 z^2+ 14 z + 13)\bigg) \right] u(p),
\end{align}
where $\Delta= (1-z)^2 m_\mu^2 + z m_G^2$.
\begin{figure}[htb]
\centering
\includegraphics[width=7cm, height=6cm]{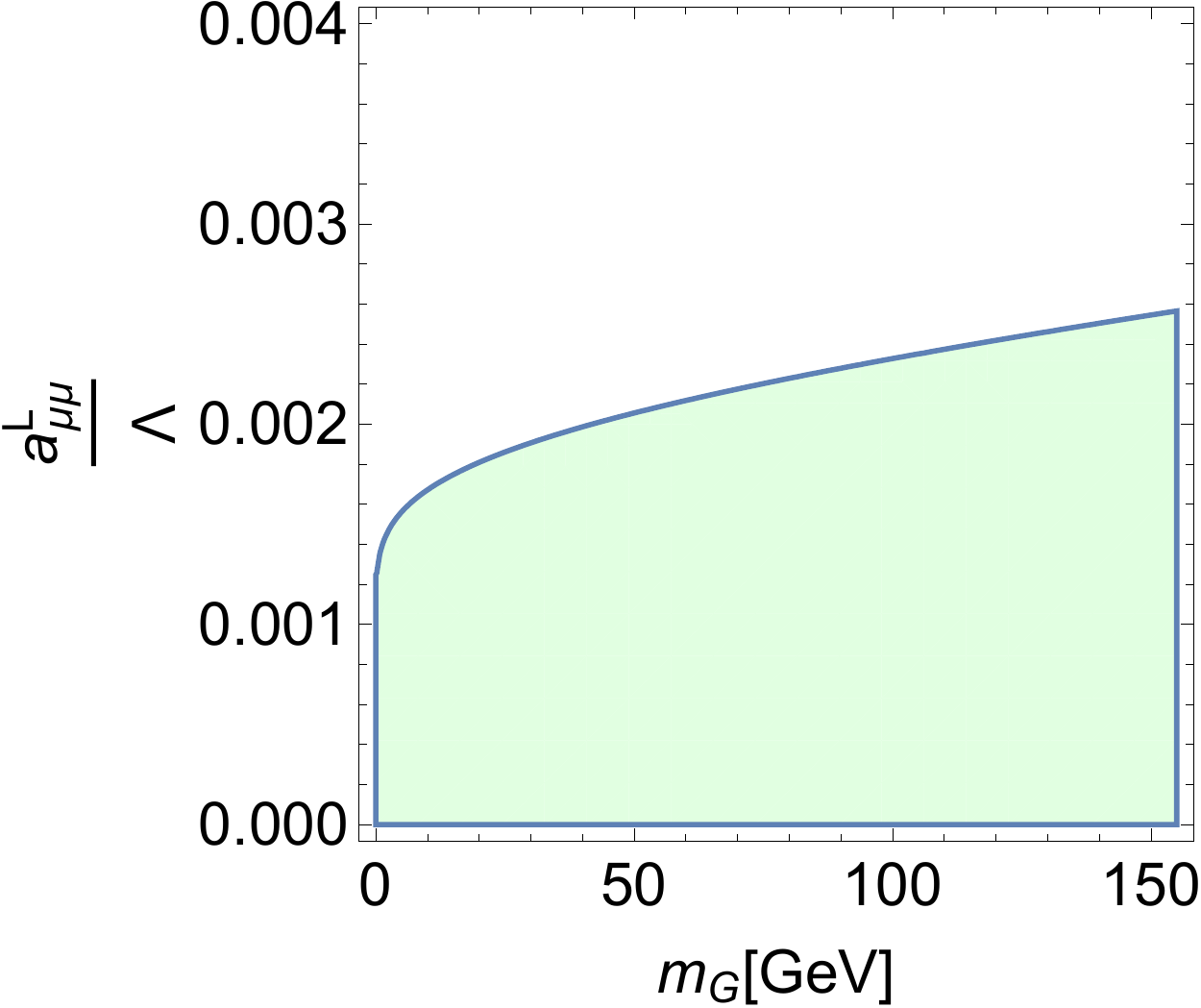}
\caption{The bound on $a^L_{\mu\mu}/\Lambda$ as a function of $m_G$ from g-2 of the muon. }
\label{fig:gm2bound}
\end{figure}
The diagram diverges and in principle one has to use full theory containing spin-2 particle, to calculate it. In order to estimate size of the $a^L_{\mu \mu}/\Lambda$ coupling, we assume  cut-off regularization and set the cut-off scale to be almost  
equal to the scale of spin-2 theory $\Lambda_N \simeq \Lambda$.  
The lower bound obtained on $a^L_{\mu\mu}/\Lambda $
from $(g-2 )_\mu$ as function of the $m_G$ is presented in Fig.~\ref{fig:gm2bound}, 
which will be used in our further calculations.
We have tried to constrain 
$a^L_{\mu \mu}$ also by considering the decay channel $Z \rightarrow 4 \mu$. This
decay channel is studied at the LHC and an upper bound is obtained for  the total decay width. The 
presence of the spin-2 particle gives additional contribution to this process  
through the $a^L_{\mu\mu}$ coupling. We checked that the constraint obtained for $a^L_{\mu\mu}$
from the $Z$ decay width to four muons is much weaker than the constraints coming from the muon anomalous magnetic moment. 
In addition to the constraints discussed above, we also consider the dilepton invariant mass distribution in $pp \rightarrow \mu^+ \mu^-$ at the 13 TeV LHC with 36.1 fb$^{-1}$ of data ~\cite{ATLAS:2017wce}.  We have performed our computation in Madgraph~\cite{Alwall:2011uj} and  find that for the parameter space considered in our model with $\Lambda \simeq$ 1000 GeV,  the LHC data in the low mass range does not constrain the scenario considered here. The result remains the same even if we allow the spin-2 mediator to couple to all the quarks with the coupling of the same order as $a^L_{sb}$.  This is mainly because the couplings of the spin-2 particle with quarks and leptons are suppressed by $\Lambda$ as opposed to the $Z'$ models which are currently being explored and constrained by the LHC dilepton data~\cite{Chivukula:2017qsi}.  

\section{The effect of spin-2 in exclusive decays $B\rightarrow K^{(*)}\mu^+\mu^-$ }

The $b \to s \mu^+ \mu^-$ transition can occur in a number of exclusive decay channels. The leptonic decay mode $B_s \to \mu^- \mu^-$ might also receive contributions from the NP. The $B_s \to \mu^- \mu^-$ amplitude arises from the contributions of axial, scalar, and pseudoscalar lepton currents as discussed in Ref. \cite{Becirevic:2012ec,Bobeth:2007dw}. In the case spin-2 mediator all relevant contributions  are either proportional to the muon mass leading to the negligible effects or kinematically  equal to zero.

We will show that higher-dimensional operators (\ref{eq:newoperators}) will produce new angular dependencies in exclusive $B \to K^{(*)} \mu^+\mu^-$ decays. 
%
The opportunity to use the measurement of higher moments of differential angular distributions to 
test physics beyond the SM coming from contributions of ${\cal O}(m_B^2/\Lambda^2)$ suppressed higher-dimensional operators and to differentiate their contribution from the estimated  higher-order QED
corrections was discussed in \cite{Zwicky}, where the authors have used the general helicity formalism to derive full angular distributions for the complete dimension-six effective Hamiltonian in those decays. 

\subsection{ The spin-2 mediator  in $B\rightarrow K\mu^+\mu^-$ } 

In our study we follow the approach and notations introduced in \cite{Becirevic:2012ec,Bobeth:2007dw} (for details see Appendix B). The angular observables  are very useful to identify NP effects in $B\rightarrow K\mu^+\mu^-$.  The angle  $\theta$ is defined as the angle between the direction of $B$ 
and the lepton $\ell^-$ in the center of mass frame of the lepton pair,  
$q^2$ is running from $q^2_{min}=4 m_\ell^2$ to $q^2_{max}=(m_B-m_K)^2$. 
As usually  we follow definitions $\beta_\ell(q^2)=\sqrt{1-4m_\ell^2/q^2}$ and 
$\lambda(q^2) = q^4+ m_B^4 + m_K^4 -2 \left( m_B^2 m_K^2 +m_B^2 q^2 + m_K^2 q^2 \right)$. 
Using this notation, we write down the double differential decay rate with respect to $q^2$ and $\cos\theta$ for the process
$B\rightarrow K\ell^+\ell^-$ as: 
\begin{eqnarray}\label{eq:tdWKll}
\frac{d^2\Gamma_\ell}{dq^2 d\cos\theta} &=& a_\ell(q^2)+
\tilde{b}_\ell(q^2) \cos\theta + c_\ell(q^2)\cos^2\theta + \tilde{d}_\ell(q^2) \cos^3\theta +\tilde{e}_{\ell}(q^2) \cos^4\theta,  
\end{eqnarray}
where we have
\begin{align}
 a_{\ell}(q^2)=& \ \mathcal{N}(q^2)
 \Big[  q^2\left(\beta_\ell^2(q^2)\lvert F_S^{NP}(q^2)\rvert^2 + \lvert F_P(q^2) \rvert^2\right) + \frac{\lambda (q^2)}{4}\left( \lvert F_A (q^2)\rvert^2 +  \lvert F_V(q^2) \rvert^2 \right)  \nonumber \\ 
 & + 4 m_{\ell}^2 m_B^2 \lvert F_A(q^2) \rvert^2+2m_{\ell}  
 \left( m_B^2 -m_K^2 +q^2\right) \text{Re}\left( F_P(q^2) F_A^{\ast}(q^2) \right)\Big]  \,,\nonumber \\ 
\tilde{b}_\ell (q^2) = &\  2 \mathcal{N}(q^2) \left[m_{\ell}  \sqrt{\lambda (q^2)} \beta_{\ell}(q^2) \text{Re} \left( F^{NP}_S(q^2) F_V^{\ast}(q^2)
 \right)+q^2 \text{Re} \left( F^{NP}_P(q^2) F_P^{\ast}(q^2)\right)\right. \nonumber \\
& \left.+\frac{\lambda (q^2)}{4}\text{Re}\left( F_A^{NP} (q^2) F_A^{\ast} (q^2)   +  F_V^{NP} (q^2) F_V^{\ast} (q^2) \right) 
+ 4 m_{\ell}^2 m_B^2 \text{Re} \left( F^{NP}_A(q^2) F_A^{\ast}(q^2)\right)\right. \nonumber \\
& \left.+ m_{\ell}  \left( m_B^2 -m_K^2 +q^2\right) \text{Re}\left( F_P^{NP}(q^2) F_A^{\ast}(q^2)+ F_P(q^2) F_A^{NP\ast}(q^2) \right)\right. \nonumber \\
&+\left. \frac{\sqrt{\lambda(q^2)} }{8}
\beta_{\ell}(q^2) f_+(q^2) C^{(8)}_{LL} \left\lbrace 4\frac{m_\ell^2}{q^2} C_{10}\left((m_B^2-m_K^2)^2 f_0(q^2)
 \right. \right.\right. \nonumber \\
&-\left.\left.\left.\lambda(q^2)f_+(q^2)\right)-  \lambda(q^2) (F_V (q^2)- F_A (q^2))\right\rbrace \right]  \,, \nonumber \\
c_\ell (q^2) = &\   \mathcal{N}(q^2)  \Big[- \frac{\lambda (q^2)}{4} \beta_{\ell}^2(q^2) 
\left( \lvert F_A(q^2) \rvert^2 +  \lvert F_V(q^2) \rvert^2 \right) + q^2 \lvert F_P^{NP}(q^2) \rvert^2+ \frac{\lambda (q^2)}{4}\left( \lvert F_A^{NP} (q^2)\rvert^2 \right. \nonumber \\
&\left. +  \lvert F_V^{NP}(q^2) \rvert^2 \right)  + 4 m_{\ell}^2 m_B^2 \lvert F_A^{NP}(q^2) \rvert^2+2m_{\ell}  
 \left( m_B^2 -m_K^2 +q^2\right) \text{Re}\left( F_P^{NP}(q^2) F_A^{NP\ast}(q^2) \right)  \nonumber \\
 &+\frac{\beta_\ell^2(q^2)}{8}f_+(q^2) \lambda(q^2) \left\lbrace C_{LL}^{(8)}\left(C_{LL}^{(8)} f_+(q^2)\left( \lambda(q^2)+2m_\ell^2 \left(2(m_B^2+m_K^2)-q^2\right)\right) \right.\right. \nonumber \\
&\left. \left. +\frac{8}{m_b} F_S^{NP}(q^2)\right)\right\rbrace
 \Big]\,, \nonumber \\
\tilde{d}_\ell(q^2)= &\mathcal{N}(q^2)\left[\frac{1}{4}
\beta_{\ell}^3(q^2) \lambda^{3/2}(q^2)f_+(q^2)C^{(8)}_{LL}
\left( F_V (q^2) -  F_A(q^2) \right) \right.\nonumber \\
&- \left. \frac{\lambda (q^2)}{4} \beta_{\ell}^2(q^2)  2\text{Re}\left( F_A^{NP} (q^2) F_A^{\ast} (q^2)   +  F_V^{NP} (q^2) F_V^{\ast} (q^2) \right)\right],\nonumber \\
\tilde{e}_\ell(q^2)= & \mathcal{N}(q^2)  \Big[- \frac{\lambda (q^2)}{4} \beta_{\ell}^2(q^2) 
\left( \lvert F_A^{NP}(q^2) \rvert^2 +  \lvert F_V^{NP}(q^2) \rvert^2 \right) -\frac{\beta_\ell^4}{8}f_+(q^2) \lambda(q^2)^2 (C_{LL}^{(8)})^2\Big]
\end{align}
with  
\begin{equation}
\mathcal{N}(q^2)=\frac{G_F^2 \alpha^2 \lvert V_{tb} V_{ts}^{\ast} \rvert^2}{512 \pi^5 m_B^3}
\beta_{\ell}(q^2)   \sqrt{\lambda (q^2) }. 
\end{equation}
Functions $\tilde{b}(q^2), \tilde{d}(q^2)$ and $\tilde{e}(q^2)$ are zero in the SM.  The terms proportional to $cos \theta$ might appear in some extensions of the SM (see e.g. \cite{Becirevic:2012ec}) while the terms $\tilde{d}(q^2)$ and $\tilde{e}(q^2)$ are genuine consequence of the spin-2 mediator. 

The SM form-factors are extended by the effects of spin-2 particle. Following  the notation of Ref. \cite{Becirevic:2012ec} we write
\begin{eqnarray}\label{F_BKell}
F_V (q^2)  &=&  C_9  f_+ (q^2) + \frac{2 m_b}{m_B +m_K} 
C_7  f_T (q^2),\quad\quad   F^{NP}_V (q^2) =  -\beta_\ell  \sqrt{\lambda (q^2) } C^{(8)}_L  f_+ (q^2)  \,,  \nonumber \\ 
F_A (q^2)    &=& C_{10}  f_+ (q^2),\quad\quad  F^{NP}_A (q^2) =  -\beta_\ell  \sqrt{\lambda (q^2) } C^{(8)}_{L5}  f_+ (q^2) \,,   \nonumber\\
 F_P (q^2)  &=& - m_{\ell}  C_{10} \left[f_+ (q^2) - \frac{m_B^2-m_K^2}{q^2}
\left(f_0  (q^2) - f_+ (q^2) \right) \right]\,,  \nonumber \\
F^{NP}_P (q^2)  &=& m_{\ell} \beta_\ell  \sqrt{\lambda (q^2) } C^{(8)}_{L5} \left[f_+ (q^2) - \frac{m_B^2-m_K^2}{q^2}
\left(f_0  (q^2) - f_+ (q^2) \right) \right]\,,  \nonumber \\
F^{NP}_S (q^2) &=&    m_\ell \frac{m_B^2 -m_K^2}{2}  \left( C^{(8)}_S+ C^{(8)}_{S'} \right) f_0 (q^2) .
\end{eqnarray}
The spin-2 particle contributions to the total decay width
are proportional to $ (C_i^{(8)})^2 \sim 1/(\Lambda^4m_G^4)$ and therefore they are insignificant.  

The interference terms proportional 
to $\cos\theta$  and $\cos^3\theta$ will however, contribute to the forward-backward asymmetry ($A_{FB}$), which is zero in the SM. The forward-backward asymmetry is defined as
\begin{eqnarray}
A_{FB}=\frac{1}{\Gamma_\ell}\int dq^2 \left(\int_{0}^1-\int_{-1}^0\right)d\cos\theta \left(\frac{d^2\Gamma_\ell}{dq^2d\cos\theta}\right)=\frac{1}{\Gamma_\ell}\int dq^2 \left(\tilde{b}_\ell(q^2)+\frac{\tilde{d}_\ell(q^2)}{2}\right), 
\end{eqnarray}
where the total decay rate is given by
\begin{eqnarray}
\label{eq:totalG}
\Gamma_\ell = 2  \int dq^2 \left(a_\ell (q^2) + \frac{c _\ell(q^2)}{3} + 
\frac{\tilde{e}_\ell(q^2)}{5} \right).
\end{eqnarray}
In our calculation we consider only region of the square of the invariant dilepton mass $1\le q^2 \le 6$ GeV$^2$. 
The odd power of $\cos\theta$  proportional to $\tilde{b}_{\ell}(q^2)$ and  $\tilde{d}_{\ell}(q^2)$, which contains the interference term
of the SM with the spin-2 contribution survives. A non-zero measurement of this asymmetry, therefore will be a clear signal of this NP scenario. 
It is interesting that spin-2  dimension-8 operators introduce dependence on ${\cos \theta}$ and ${\cos^3 \theta}$, not present in the SM. 
\begin{figure}
\centering
\begin{minipage}{0.45\linewidth}
\centering
\includegraphics[width=7.0cm, height=5.3cm]{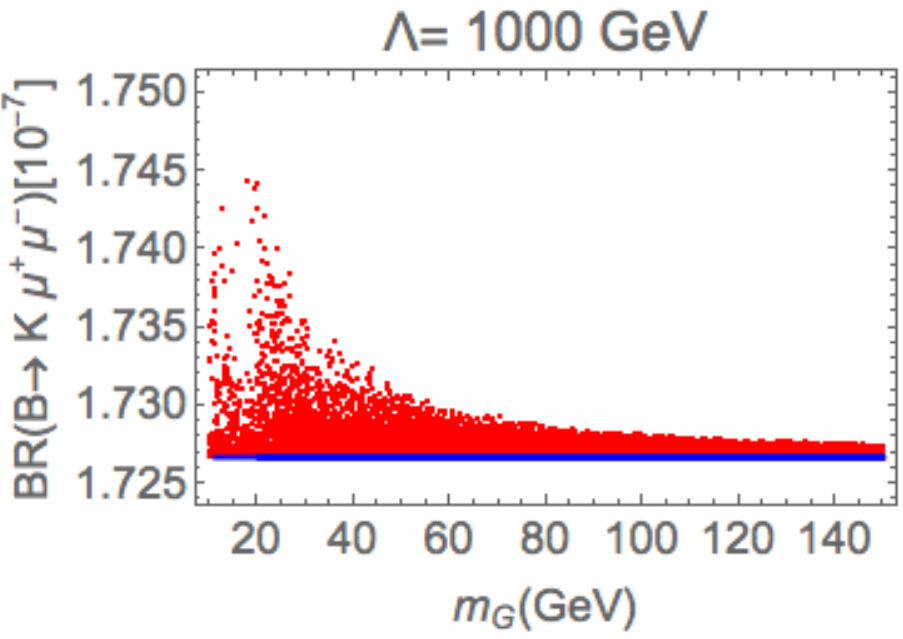}
\caption{Br($B\rightarrow K\mu^+\mu^-$) as a function of spin-2 mass $m_G$  for $\Lambda$ =  1000 GeV in the presence of the spin-2 contribution [red] taking into account the constraints from the muon anomalous magnetic moment, $B_s-\bar{B}_s$ mixing and the top decay width. The SM value for the BR is shown by a blue line.}
\label{fig:br}
\end{minipage}
\hspace{1cm}
\begin{minipage}{0.45\linewidth}
 \includegraphics[width=7.0cm, height=5.3cm]{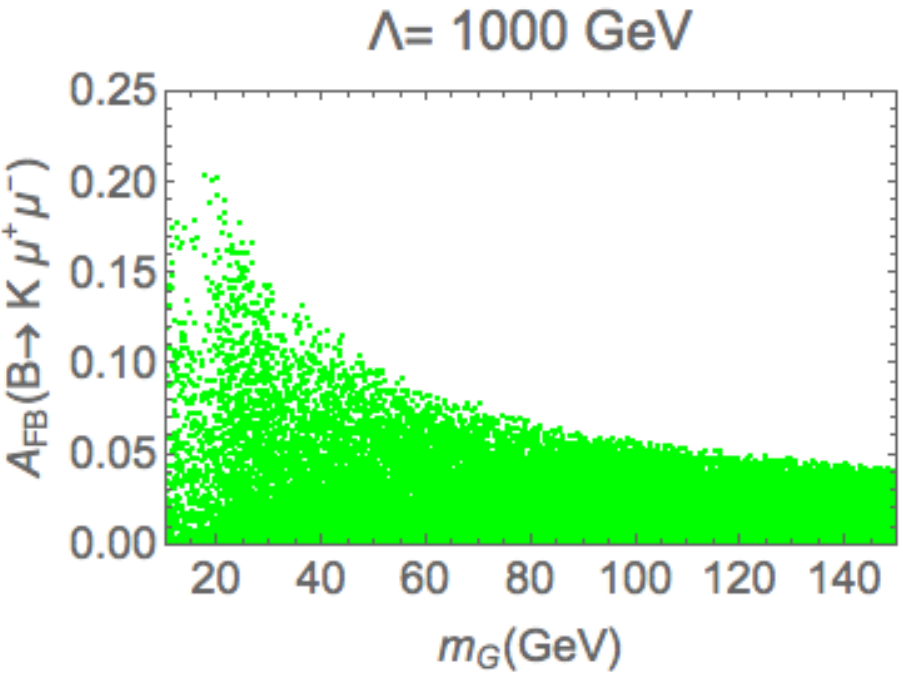}
 \caption{The forward-backward asymmetry $A_{FB}$ as a function of spin-2 mass $m_G$, taking into account the  constraints from the $B_s-\bar{B}_s$ mixing, the top decay width and muon anomalous magnetic moment for $\Lambda$ = 1000 GeV in the presence of spin-2 contribution. Note that the SM value of $A_{FB}$ is zero.}
 \label{fig:fb}
\end{minipage}
\end{figure}
The terms $a_{\ell}(q^2)$, $c_{\ell}(q^2)$ and  $\tilde{e}_{\ell}(q^2)$ contributing to the 
total decay width  is sensitive to the spin-2  contribution in the form of NP$^2 \approx 1/(16\Lambda^2m_G^2)^2$. 
This in turn leads for the increase of the order of $\sim 1-2\%$ as presented in Fig. \ref{fig:br}. 
Our hypothesis on the muon specific interaction of the spin-2 particle implies that the first lepton generation does not receive the spin-2 corrections. However, due to the tiny increase of the differential branching ratio $BR(K\rightarrow K \mu^+ \mu^-)$ the ratio $R_K$ is only insignificantly different from $1$. 
Obviously, the interaction of the spin-2 mediator with the $b$ and $s$ quarks as well as muons, without additional new particles, like vector-like fermions or gauge bosons cannot explain $R_{K^{(*)}}$ anomaly. We present  in Fig.~\ref{fig:fb} $A_{FB}$ as a function of $m_G$, for all the points satisfying the bounds coming from the  $B_s-\bar{B}_s$ mixing, the top decay width and the muon anomalous magnetic moment.  For the rather low values of the spin-2 particle mass  $m_G\simeq 20 - 40$ GeV,  the generated forward-backward can be as large as  $20 \%$. 

There is $\cos^4\theta$ dependence as a spin-2 particle effect in Eq.(\ref{eq:tdWKll}), proportional to $\tilde{e}_\ell(q^2)$. We find that $E_\ell = (1/\Gamma_\ell)\int dq^2 \tilde{e}_\ell(q^2)$, being a contribution of the order $(1/\Lambda)^4$, is insignificant  and can be only $\sim 1\%$ of the leading contribution coming from the term proportional to $a_l(q^2)$,  for $m_G$ in the range 20$-$50 GeV and $\Lambda$ = 1000 GeV.  Another observable considered in \cite{Bobeth:2007dw,experim} is $F^\ell_H = (2/\Gamma_\ell)\int dq^2 \left(a_\ell(q^2)+c_\ell(q^2)\right)$. This  term being proportional to $m_\mu^2$ is small in the SM. The inclusion of the  spin-2 contributions modifies negligibly this observable by the terms proportional to $(1/\Lambda)^4$ .

It is worth pointing out that the left-handed nature of the muonic coupling to spin-2 particle will induce couplings of spin-2 particle to neutrinos. That will imply that the branching ratio of $B\to K^{(*)} \nu \nu$ might obtain additional contributions.
The Belle collaboration  determined the upper bounds for the branching fractions of these decays decays \cite{Grygier:2017tzo}.  
The effect of spin-2 particle  in  $BR(B\to K^{(*)} \nu \bar\nu) $ is  almost equal to the effect of spin-2 in  the branching ratio for 
 $B\rightarrow K^{(*)}\mu^+\mu^-$. Therefore, it   contributes  insignificantly  to the SM prediction.

\subsection{Spin-2  mediator in the $B \rightarrow K^{\ast}(\rightarrow K \pi)\mu^+\mu^-$ decay} 
 The  spin-2 particle can also affect the process $B \rightarrow K^{\ast}(\rightarrow K \pi)\mu^+\mu^-$.  The relevant form factors
 for the $B\rightarrow K^\ast$  matrix elements are defined in Ref. \cite{Altmannshofer:2008dz,Kruger:2005ep} (for details see Appendix B), 
with $q^\mu=p^\mu-p_{K^*}^\mu$ and $\epsilon_\mu$ as the polarization vector of $K^\ast$. 
The angular distribution of the four  body decay  $B \rightarrow K^{\ast}(\rightarrow K \pi)\ell^+\ell^-$
has been discussed extensively in  the literature (see e.g. Ref.   \cite{Altmannshofer:2008dz,Kruger:2005ep}). We determine contributions of the operators ${\mathcal O}^{(8)}_{L}$, ${\cal O}^{(8)}_{L5}$ and  ${\mathcal O}^{(8)}_{LL}$ to the 
$B \rightarrow K^{\ast}\mu^+\mu^-$decay amplitude
\begin{equation}
  \frac{d^4\Gamma}{dq^2\, d\cos\theta_\ell\, d\cos\theta_{K^*}\, d\phi} =
   \frac{9}{32\pi} I(q^2, \theta_\ell, \theta_{K^*}, \phi)\,,
\end{equation}
where
\begin{align}\label{eq:bKll_angdist}
  I(q^2, \theta_\ell, \theta_{K^*}, \phi)& = 
      I_1^s \sin^2\theta_{K^*}+ I_1^c \cos^2\theta_{K^*}
      + (I_2^s \sin^2\theta_{K^*}+ I_2^c \cos^2\theta_{K^*}) \cos 2\theta_\ell,
\nonumber \\       
    & + I_3 \sin^2\theta_{K^*} \sin^2\theta_\ell \cos 2\phi 
      + I_4 \sin 2\theta_{K^*} \sin 2\theta_\ell\cos\phi 
\nonumber \\       
    & + I_5 \sin 2\theta_{K^*} \sin\theta_\ell \cos\phi
\nonumber \\      
    & + (I_6^s \sin^2\theta_{K^*} +
      {I_6^c \cos^2\theta_{K^*}})  \cos\theta_\ell
      + I_7 \sin 2\theta_{K^*} \sin\theta_\ell \sin\phi
\nonumber \\ 
    & + I_8 \sin 2\theta_{K^*} \sin 2\theta_\ell\sin\phi
      + I_9 \sin^2\theta_{K^*} \sin^2\theta_\ell \sin 2\phi\,.
\end{align}
The angular coefficients $I_i$'s in the presence of standard dimension-6 operators are dependent on $q^2$. The full expressions for $I_i$ as a functions of the  transversity amplitudes  are given in Appendix B. 
 We define $\theta_{K^*}$ as the angle of direction of flight of $K$ 
in the $K^*$ rest frame with respect to the direction of flight of $K^*$ in the $B$ rest frame,  $\theta_{\ell}$ as the angle of direction of flight of $\ell^-$ in the dilepton rest frame with respect to the direction of flight of the two leptons in the $B$ rest frame . The angle between the plane formed by $K\pi$ and $\ell\ell$ is denoted by $\phi$. 

\begin{figure}
\includegraphics[width=7.0cm, height=6cm]{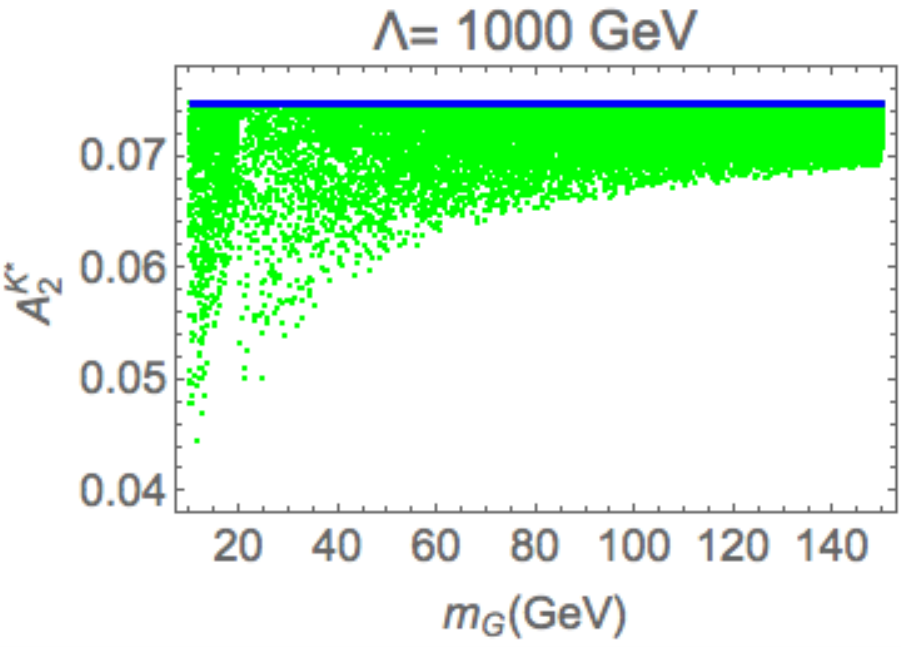}
\hspace{1.0cm}
\includegraphics[width=7.0cm, height=6cm]{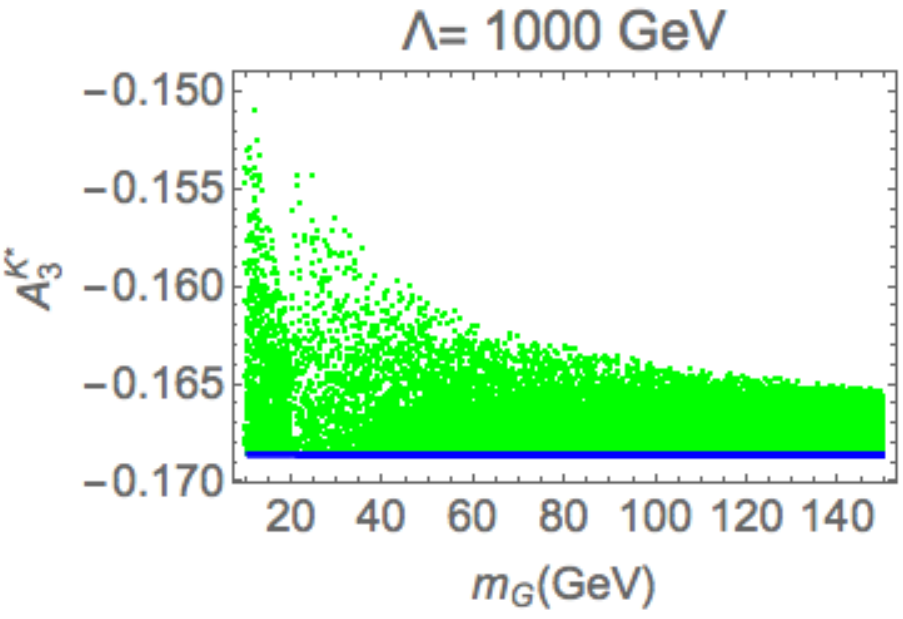}
\caption{$A_{2}^{K^*},A_{3}^{K^*}$ as a function of the spin-2 mass, taking into account the  constraints coming from the $B_s-\bar{B}_s$ mixing, the top decay width and muon anomalous magnetic moment for the spin 2 contribution [green] for $\Lambda$ = 1000 GeV. The SM value for the $A_{2,3}^{K^*}$ is shown by a blue line.}
\label{fig:kstarasym}
\end{figure}

The angular coefficients are combinations of the transversity 
amplitudes $A_{\perp,\parallel,0,t}^{L,R}$ which are listed in Ref.\cite{Altmannshofer:2008dz}.  We have also followed the convention of   \cite{Altmannshofer:2008dz} for the polarization vector of the virtual gauge boson $\epsilon_{V^*}^\mu(n)$, $n=\pm$ (transverse),  $n=0$ (longitudinal) or $n=t$ (timelike) and the $K^*$ 
polarization vector $\epsilon_{K^*}^\mu(m)$, with $m=\pm,0$. 
The  transversity amplitudes $A_{\perp,\parallel,0,t}^{L,R}$ 
in our case get modified by the dimension 8 operators and are listed below.
\begin{align}
A_{\perp L,R}  &=  \mathcal{N}_{K^*}(q^2) \sqrt{2} \lambda_{K^*}^{1/2} \bigg[ 
\left[ (C_{9} -\mathcal{B}_{K^*} C^{(8)}_L) \mp (C_{10} -\mathcal{B}_{K^*}  C^{(8)}_{L5}) \right] \frac{ V(q^2) }{ m_B + m_{K^*}} 
 + \frac{2m_b}{q^2} C_7  T_1(q^2) \bigg],\nonumber \\
A_{\parallel L,R}  & = - \mathcal{N}_{K^*}(q^2) \sqrt{2}(m_B^2 - m_{K^*}^2) \bigg[ \left[ (C_9 -\mathcal{B}_{K^*}  C^{(8)}_L) \mp (C_{10}-\mathcal{B}_{K^*}  C^{(8)}_{L5}) \right] 
\frac{A_1(q^2)}{m_B-m_{K^*}} \nonumber \\
& +\frac{2m_b}{q^2} C_7  T_2(q^2)\bigg], \nonumber \\
A_{0L,R}  &=  - \frac{\mathcal{N}_{K^*}(q^2)}{2 m_{K^*} \sqrt{q^2}}  \bigg\{ 
 \left[ (C_9-\mathcal{B}_{K^*} C^{(8)}_L) \mp (C_{10}-\mathcal{B}_{K^*} C^{(8)}_{L5} ) \right]
\nonumber\\
 & \qquad \times 
\bigg[ (m_B^2 - m_{K^*}^2 - q^2) ( m_B + m_{K^*}) A_1(q^2) 
 -\lambda_{K^*} \frac{A_2(q^2)}{m_B + m_{K^*}}
\bigg] 
\nonumber\\
& \qquad + {2 m_b}C_7\bigg[
 (m_B^2 + 3 m_{K^*}^2 - q^2) T_2(q^2)
-\frac{\lambda_{K^*}}{m_B^2 - m_{K^*}^2} T_3(q^2) \bigg]\bigg\} +A^{NP}_{0L,R}
,\nonumber\\
 A_t  &= \frac{\mathcal{N}_{K^*}(q^2)}{\sqrt{q^2}}\left\lbrace \lambda_{K^*}^{1/2} \left[ 2 (C_{10}-\mathcal{B}_{K^*}  C^{(8)}_{L5})  \right] A_0(q^2)-\beta_\ell(q^2) \cos\theta_\ell
 \frac{(m_B- m_{K^*})}{m_{K^*}}\right. \nonumber \\
 &\quad\left. \times \left[ (m_B^2 - m_{K^*}^2 - q^2) ( m_B + m_{K^*})^2 A_1(q^2) - \lambda_{K^*}  A_2(q^2) \right] C^{(8)}_{LL}\right\rbrace, \nonumber  \\
A_S  &= - 2\mathcal{N}_{K^*}(q^2) \lambda^{1/2}_{K^*} m_b m_\ell (C^{(8)}_{S} - C^{(8)}_{S'})  A_0(q^2), 
\end{align}
where
$\mathcal{B}_{K^*} = \beta_\ell \sqrt{\lambda_{K^*}}\cos\theta_\ell $, with $\lambda_{K^*} = m_B^4  + m_{K^*}^4 + q^4 - 2 (m_B^2 m_{K^*}^2+ m_{K^*}^2 q^2  + m_B^2 q^2)$ 
\begin{eqnarray}
\mathcal{N}_{K^*}(q^2)&=& V_{tb}^{\vphantom{*}}V_{ts}^* \left[\frac{G_F^2 \alpha^2}{3\cdot 2^{10}\pi^5 m_B^3}
 q^2 \lambda^{1/2}_{K^*}
\beta_\ell(q^2) \right]^{1/2}, \nonumber 
\end{eqnarray}
and
\begin{eqnarray}
A^{NP}_{0L} &=& \frac{\mathcal{N}_{K^*}(q^2)}{m_{K^*} } \sqrt{\frac{\lambda_{K^*}}{q^2}} \beta_\ell(q^2)\cos\theta_\ell \bigg[(m_B^2 - m_{K^*}^2 - q^2) ( m_B + m_{K^*}) A_1(q^2)  \nonumber \\
&&-\lambda_{K^*} \frac{A_2(q^2)}{m_B + m_{K^*}}\bigg]  C^{(8)}_{LL}, \nonumber \\
A^{NP}_{0R} &=& 0 .
\end{eqnarray}
We point out that Wilson coefficients $C^{(8)}_{L,L5}$ and $C^{(8)}_{LL}$ in our case are always accompanied by $\cos\theta_\ell $.  In the limit of massless leptons and 
considering only the interference of $C^{(8)}_{L,L5}, C^{(8)}_{LL}$ with $C_{7,9,10}$ (i.e. neglecting higher powers of $C^{(8)}_i$ coefficients), all the angular functions 
$I_i$'s are of the form $I_i^{SM}+ C_{7,9,10} (C^{(8)}_{L,L5}, C^{(8)}_{LL}) \cos\theta_\ell \approx I_i^{SM}+I_i^{int} \cos\theta_\ell $. The $I_i^{SM}$ are the values of the angular coefficients  in 
the SM, whereas $I_i^{int}$ denotes the interference term of spin-2 dimension-8 operators with the standard dimension-6 ones.  
Such dependence on 
 $\cos\theta_{\ell}$  
of the NP terms leads us to consider the forward-backward asymmetry in $\theta_{\ell}$, 
\begin{align}
A_{FB}^{K^*}&=\frac{1}{\Gamma}\int dq^2\int_0^{2\pi}d\phi \int_{-1}^{1}d\cos\theta_{K^*} \left(\int_{0}^{1}-\int_{-1}^{0}\right) d\cos\theta_\ell \left(\frac{d^4\Gamma}{dq^2\, d\cos\theta_\ell\, d\cos\theta_{K^*}\, d\phi} \right) \nonumber \\
&=\frac{3}{8}\left(2 I_6^{s(SM)}+I_6^{c(int)}+ I_1^{c(int)}+2I_1^{s(int)}\right).
\end{align}
\begin{figure}
\centering
\begin{minipage}{0.45\linewidth}
\centering
\includegraphics[width=7.0cm, height=5.3cm]{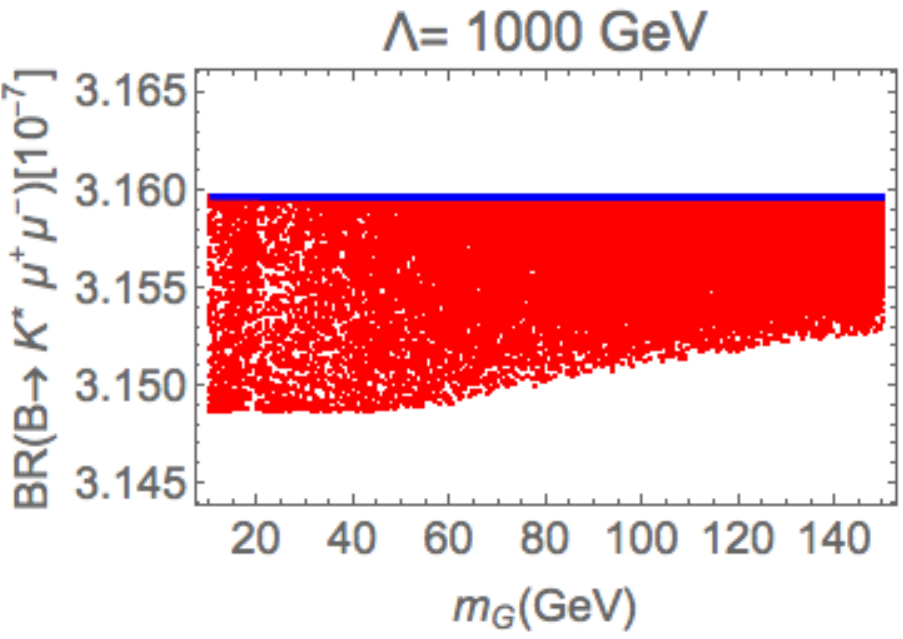}
\caption{Br($B\rightarrow K^*\mu^+\mu^-$)  at $1\le q^2 \le 6$ GeV$^2$ as a function of $m_G$ for $\Lambda$ = 1000 GeV in the presence of the spin-2 contribution [red] taking into account the constraints from the muon anomalous magnetic moment, $B_s-\bar{B}_s$ mixing and the top decay width. The SM value for the BR is shown by a blue line.}
 \label{fig:brKstar}
\end{minipage}
\hspace{1.0cm}
\begin{minipage}{0.45\linewidth}
 \includegraphics[width=7.0cm, height=5.3cm]{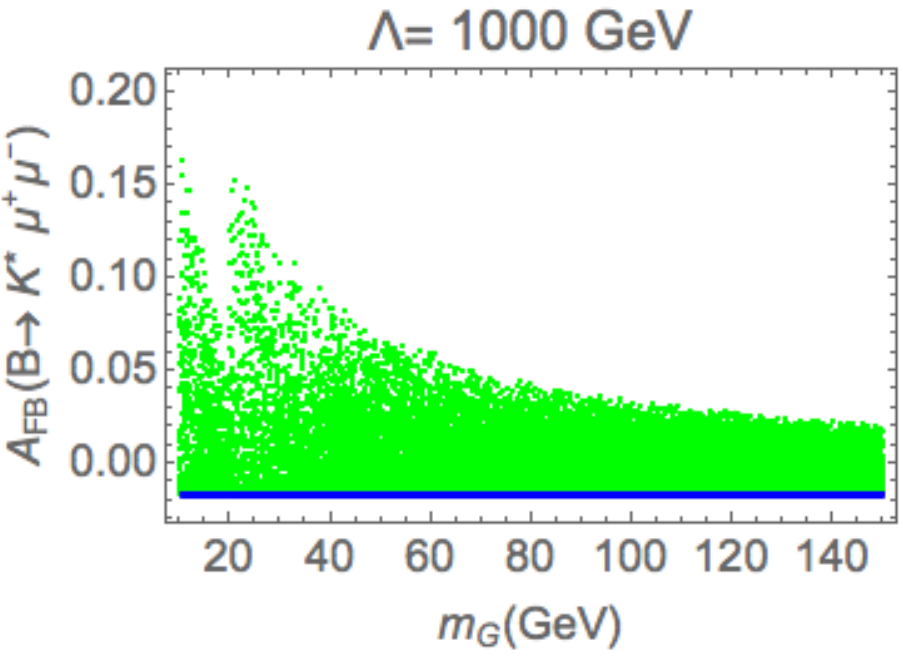}
 \caption{$A_{FB}$ as a function of the spin-2 mass, taking into account the  constraints coming from the $B_s-\bar{B}_s$ mixing, the top decay width and muon anomalous magnetic moment for the spin 2 contribution [green] for $\Lambda$ = 1000 GeV. The SM value for the $A_{FB}$ is shown by a blue line. }
 \label{fig:fbKstar}
\end{minipage}
\end{figure}
\noindent 
We present in Fig.~\ref{fig:brKstar} the branching ratio for $B\rightarrow K^*\mu^+\mu^-$ at low-energy region $1\le q^2 \le 6$ GeV$^2$ and in Fig.~\ref{fig:fbKstar}  the forward-backward asymmetry, as a function of the mass of spin-2 particle, respecting all relevant constraints from the $B_s - \bar B_s$ mixing, the top decay width and muon anomalous magnetic moment. The spin-2 mediator increases  the forward-backward asymmetry for the decay $B\rightarrow K^* \mu^+ \mu^-$ relative to the SM value $A_{FB,SM}^{K^*}=-0.035^{+0.036}_{-0.033}$~\cite{Descotes-Genon:2013vna}.
For the rather low values of the spin-2 particle  mass $m_G\simeq 20 -4 0$ GeV,  the generated forward-backward can be close to  $ 15\%$.   As in the case of $B\to K \mu^+ \mu^-$, due to the insignificant increase of the  branching ratio for $B\rightarrow K^*\mu^+\mu^-$, inclusion of the spin-2  mediator cannot explain the $R_{K^{(*)}}$ problem. The spin-2 particle also modifies the triple forward-backward asymmetries for $K^*$ and  the leptons and also the double asymmetries in $\phi$ and $\theta_K$: 
\begin{align}
A_{2}^{K^*}&=\frac{1}{\Gamma}\int dq^2\left(\int_{-\pi/2}^{\pi/2}- \int_{\pi/2}^{3\pi/2}\right)d\phi \left(\int_{0}^{1}-\int_{-1}^{0}\right)d\cos\theta_{K^*} \left(\int_{0}^{1}-\int_{-1}^{0}\right) d\cos\theta_\ell ~ \Gamma_{tot}\nonumber \\
&=\frac{1}{\pi}\left(2I_4^{(SM)}+ I_5^{(int)}\right), \\
A_{3}^{K^*}&=\frac{1}{\Gamma}\int dq^2\left(\int_{-\pi/2}^{\pi/2}- \int_{\pi/2}^{3\pi/2}\right)d\phi \left(\int_{0}^{1}-\int_{-1}^{0}\right)d\cos\theta_{K^*} \int_{-1}^{1} d\cos\theta_\ell ~ \Gamma_{tot}\nonumber \\
&=\frac{3}{8}\left(2I_5^{(SM)}+ I_4^{(int)}\right), 
\end{align}
where $\Gamma_{tot}$= $d^4\Gamma/\left(dq^2\, d\cos\theta_\ell\, d\cos\theta_{K^*}\, d\phi\right)$. We show in Fig.~\ref{fig:kstarasym} the asymmetries $A_{2}^{K^*},A_{3}^{K^*}$ as a function of the spin-2 mass by taking into account all the experimental constraints considered in the text.

\section{Conclusions}

We have considered effects of spin-2 mediator in $b \rightarrow s \mu^+ \mu^-$ transitions.  We assumed that only flavour non-universal couplings to left-handed $b$ and $s$ quarks exist and that in the leptonic sector the spin-2 particle couples to muons only. Naturally, the $B_s - \bar B_s$ mixing and the muon anomalous magnetic moment enable us to constrain these couplings. 
It is interesting that these constraints are very severe not allowing any significant effect in the observables for the $B\rightarrow K \mu^+ \mu^-$ and $B\rightarrow K^* \mu^+ \mu^-$ decays.  The model contains derivative couplings of the spin-2 particle with fermions.  
The dimension eight effective Lagrangian induced by the spin-2 mediator creates additional terms in the angular distributions of the $B\rightarrow K^{(*)} \mu^+ \mu^-$ decays. Such contributions are of the order $1/\Lambda^4$ and therefore negligible. 
However we find out that the spin-2 effective Lagrangian can create forward-backward asymmetries of the order $\sim 10-20 \%$,  for rather small mass of the spin-2 particle  $m_G\simeq 20 - 40$ GeV.  This is specially interesting for $B  \rightarrow K \mu^+ \mu^-$ decay where such an asymmetry vanishes in the SM.

Due to the left-handed nature of $b$, $s$ and $\mu$ in  our model,  one can also generate the coupling of spin-2 particle to the top and charm quark $a^L_{bs} \simeq a^L_{tc}$.  In that case the spin-2 particle can be seen at the LHC,  it would mean that for its mass higher than the top quark mass, the possible decay channels would be the top quark and charm quark jet, the $b$ jet and light quark jet, as well as muon pair.  A dedicated study of spin-2 particle in the process $b \bar s \to \mu \mu$ as performed in Ref. \cite{Greljo:2017vvb}  would put  additional constraints on the couplings discussed in this paper.

The spin-2 mediation increases  insignificantly  the branching ratios  $BR(B\rightarrow K^{(*)} \mu^+ \mu^-)$, while the branching ratios $BR(B\rightarrow K^{(*)} e^+ e^-)$ are the SM-like. That implies that  our simple proposal  can not explain the observed $R_K$ and $R_{K^*}$ anomalies.  One might expect that some full fledged model containing additional scalars or gauge bosons together with spin-2 particle can account observed anomalies and also generate the  forward-backward asymmetry.

{\bf Acknowledgment:}  We owe a debt of gratitude to Christoph Bobeth for an illuminating discussion on the dimension eight effective Lagrangian and for a careful reading of the  manuscript. 
We are grateful to Roman Zwicky and D. Be\v cirevi\'c for useful discussions. 
The work of B.M. and M.P. is funded by the European Union's Horizon 2020 research and
innovation program under the Twinning grant agreement No. 692194, RBI-T-WINNING.
The  work  of B. M. and M.P. is also supported by the Croatian Science Foundation (HRZZ) project No. 5169,
PhySMaB. S.F.  and M.P. acknowledge support of the Slovenian Research Agency through research core funding No. P1-0035.

\appendix

\section{Matrix elements of the operators in the $B_s -\bar{B}_s$ mass difference} 

 The operators contributing  to the mass difference in Eq. (\ref{MBs}) can be determined using the results of QCD running given in Ref.  \cite{Becirevic:2001jj}: 
\begin{eqnarray}
(\Delta M_s)_{m_b^2\mathcal{Q}_1} &=& \frac{2}{3}m_{B_s}f_{B_s}^2 B_1(m_b) m_b^2(m_b) \left(\frac{(a^L_{sb})^2}{4\Lambda^2m_G^2}\right)\left[0.865-0.017\eta \right]\eta^{0.286}, \nonumber \\
(\Delta M_s)_{m_b^2\tilde{\mathcal{Q}}_2}&=& \frac{5}{12}\left(\frac{m_{B_s}}{m_b(m_b)+m_s(m_b)}\right)^2 m_{B_s}f_{B_s}^2 B_2(m_b) m_b^2(m_b)\left( \frac{(a^L_{sb})^2}{16\Lambda^2m_G^2}\frac{8}{3}\right)\nonumber \\
&&\left\lbrace\left[(1.879-0.18\eta)\eta^{-0.692}+(0.012-0.003\eta)\eta^{0.787}\right]\right.\nonumber \\
&&\left.-\left[(-0.493-0.014 \eta)\eta^{-0.692}+(0.18+0.008\eta)\eta^{0.787}\right]\right\rbrace, \nonumber \\
(\Delta M_s)_{m_b^2\tilde{\mathcal{Q}}_3}&=& -\frac{1}{12}\left(\frac{m_{B_s}}{m_b(m_b)+m_s(m_b)}\right)^2 m_{B_s}f_{B_s}^2 B_3(m_b) m_b^2(m_b)\left( \frac{(a^L_{sb})^2}{16\Lambda^2m_G^2}\frac{8}{3}\right)\nonumber \\
&&\left\lbrace\left[(-0.044+0.005\eta)\eta^{-0.692}+(0.035-0.012\eta)\eta^{0.787}\right]\right.\nonumber \\
&&\left.-\left[(0.011+0.0 \eta)\eta^{-0.692}+(0.54+0.028\eta)\eta^{0.787}\right]\right\rbrace, 
\end{eqnarray}
with $\eta = \alpha_S(\Lambda)/\alpha_S(m_t)$.

\section{$B\to K\mu^+ \mu^-$  and $B\to K^* \mu^+ \mu^-$ decays}  

In the calculation we use the following kinematics of $b\rightarrow s \mu^+ \mu^-$ decay . The lepton pair is taken to be at rest, $\bf{k_+}+\bf{k_-}$ = 0 meaning that   
\begin{eqnarray}
&& k_++k_- = (q,0,0,0),~k_-=\frac{q}{2}(1,\beta_\ell \sin\theta_\ell,0,\beta_\ell \cos\theta_\ell),
 ~k_+=\frac{q}{2}(1,-\beta_\ell \sin\theta_\ell,0,-\beta_\ell \cos\theta_\ell),\nonumber \\
&& \hspace*{-0.7cm}  {\rm and} \nonumber \\
&&p_b = (E_b,0,0,\vec{p}_b),~p_s=(E_s,0,0,\vec{p}_s),~\vec{p}_b=\vec{p}_s=\frac{1}{2q}
\sqrt{(m_b^2-m_s^2+q^2)^2-4 q^2 m_b^2)} \,,\nonumber \\
&&E_b = \frac{1}{2q}(m_b^2+q^2-m_s^2),~E_s = \frac{1}{2q}(m_b^2-m_s^2-q^2),
~~\beta_\ell=\sqrt{1-4\frac{m_\ell^2}{q^2}},
\end{eqnarray}
 where $\theta_\ell$ is the angle between the directions of $b$ and $\ell^-$.

 
The form factors for the $B \to K^*$ weak transitions are defined as:
\begin{eqnarray}
&&\braket{K^*(p_{K^*}) \left\vert \bar{s} \gamma_{\mu}P_{L,R} b \right\vert B(p)}
= i\epsilon_{\mu\nu\alpha\beta} \epsilon^{\nu*}p^{\alpha}q^{\beta} \frac{V(q^2)}{m_B+m_{K^*}}
\mp\frac{1}{2} \bigg \{\epsilon_{\mu}^*(m_B+ m_{K^*})A_1(q^2) \nonumber \\
&&  -(\epsilon^*\cdot q)(2p -q)_\mu \frac{A_2(q^2)}{m_B+m_{K^*}} -
\frac{2m_{K^*}}{q^2} (\epsilon^*\cdot q) q_\mu [A_3(q^2)- A_0(q^2)] \bigg \},
\end{eqnarray}
where 
\begin{eqnarray}
A_3 (s) = \frac{m_B+m_{K^*}}{2m_{K^*}} A_1(q^2) - \frac{m_B-m_{K^*}}{2m_{K^*}}A_2(q^2),
\end{eqnarray}
and 
\begin{eqnarray}
\lefteqn{\langle K^*(p_{K^*})|{\bar{s}i \sigma_{\mu\nu}q^{\nu}P_{R,L} b}{|{B}(p)\rangle}
= - i\epsilon_{\mu\nu\alpha\beta} \epsilon^{\nu*}p^{\alpha}q^{\beta}T_1(q^2)
\pm\frac{1}{2}\bigg\{[\epsilon_{\mu}^*(m_B^2-m_{K^*}^2)}\nonumber \\
&-& (\epsilon^*\cdot q)(2p-q)_{\mu}] T_2(q^2)
+ (\epsilon^*\cdot q)
\bigg[ q_\mu - \frac{q^2}{m_B^2-m_{K^*}^2} (2p - q)_\mu\bigg]T_3(q^2)\bigg\}.
\end{eqnarray}

In the angular distribution given in Eq.~(\ref{eq:bKll_angdist}) following combinations of the amplitudes are introduced:

\begin{align}
  I_1^s & = \frac{(2+\beta_\ell^2)}{4} \left[|{A_\perp^L}|^2 + |{A_\parallel^L}|^2 + (L\to R) \right] 
            + \frac{4 m_\ell^2}{q^2} \text{Re}\left({A_\perp^L}^{}{A_\perp^R}^* + {A_\parallel^L}^{}{A_\parallel^R}^*\right),  \nonumber 
\\
  I_1^c & =  |{A_0^L}|^2 +|{A_0^R}|^2  + \frac{4m_\ell^2}{q^2} 
               \left[|A_t|^2 + 2\text{Re}({A_0^L}^{}{A_0^R}^*) \right] + \beta_\ell^2 |A_S|^2 ,  \nonumber 
\\
  I_2^s & = \frac{ \beta_\ell^2}{4}\left[ |{A_\perp^L}|^2+ |{A_\parallel^L}|^2 + (L\to R)\right],   \quad   I_2^c  = - \beta_\ell^2\left[|{A_0^L}|^2 + (L\to R)\right],  \nonumber 
\\
  I_3 & = \frac{1}{2}\beta_\ell^2\left[ |{A_\perp^L}|^2 - |{A_\parallel^L}|^2  + (L\to R)\right],  \quad  I_4  = \frac{1}{\sqrt{2}}\beta_\ell^2\left[\text{Re} ({A_0^L}^{}{A_\parallel^L}^*) + (L\to R)\right],  \nonumber 
\\
  I_5 & = \sqrt{2}\beta_\ell\left[\text{Re}({A_0^L}^{}{A_\perp^L}^*) - (L\to R) 
- \frac{m_\ell}{\sqrt{q^2}}\, \text{Re}({A_\parallel^L} {A_S^*}+{A_\parallel^R} {A_S^*})
\right],  \nonumber 
\\
  I_6^s  & = 2\beta_\ell\left[\text{Re} ({A_\parallel^L}^{}{A_\perp^L}^*) - (L\to R) \right],  \quad   I_6^c    =
 4 \beta_\ell  \frac{m_\ell}{\sqrt{q^2}}\, \text{Re} \left[ {A_0^L} {A_S^*} + (L\to R) \right],  \nonumber 
\\
  I_7 & = \sqrt{2} \beta_\ell \left[\text{Im} ({A_0^L}^{}{A_\parallel^L}^*) - (L\to R) 
+ \frac{m_\ell}{\sqrt{q^2}}\, {\text{Im}}({A_\perp^L} {A_S^*}+{A_\perp^R} {A_S^*})
\right],  \nonumber 
\\
  I_8 & = \frac{1}{\sqrt{2}}\beta_\ell^2\left[\text{Im}({A_0^L}^{}{A_\perp^L}^*) + (L\to R)\right],  \quad  I_9  = \beta_\ell^2\left[\text{Im} ({A_\parallel^L}^{*}{A_\perp^L}) + (L\to R)\right].
\end{align}

\end{document}